\documentclass[a4paper,11pt]{article}
\pdfoutput=1 

\usepackage{jcappub} 

\usepackage[T1]{fontenc} 

\usepackage{hyperref}

\title{Towards detecting super-GeV dark matter via annihilation to neutrinos}

\author[a]{L.~Salvador Miranda,}
\author[b]{S.~Basegmez du Pree,}
\author[a]{K.~C.~Y.~Ng,}
\author[c]{A.~Cheek,}
\author[d]{C.~Arina}

\affiliation[a]{Department of Physics, The Chinese University of Hong Kong, Shatin, Hong Kong China}
\affiliation[b]{Nikhef, National Institute for Subatomic Physics, Science Park 105
1098 XG, Amsterdam, The Netherlands}
\affiliation[c]{Astrocent, Nicolaus Copernicus Astronomical Center of the Polish Academy of Sciences, ul.Rektorska 4, 00-614 Warsaw, Poland}
\affiliation[d]{Centre for Cosmology, Particle Physics and Phenomenology, Universit\'e catholique de Louvain, Louvain-la-Neuve B-1348, Belgium}

\emailAdd{luissalvadormirandapalacios@cuhk.edu.hk}
\emailAdd{s.basegmez.du.pree@nikhef.nl}
\emailAdd{kcyng@cuhk.edu.hk}
\emailAdd{acheek@camk.edu.pl}
\emailAdd{chiara.arina@uclouvain.be}

\abstract{
The next generation of neutrino telescopes will feature unprecedented sensitivities in the detection of neutrinos. Here we study the capabilities of a large-scale neutrino telescope, like the fully-operating KM3NeT experiment in the near future, for detecting dark matter annihilation signals from the Galactic Centre. 
We consider both ORCA and ARCA detectors, covering dark matter masses from a few GeV to 100 TeV. We obtain the sensitivities with a maximum-likelihood analysis method and present them as upper limits in the thermally averaged annihilation cross-section into Standard Model fermions. Our projections show that the sensitivity of such a neutrino telescope can reach the thermal relic line for $m_{\chi}\gtrsim 1\,{\rm TeV}$ and for $m_\chi \simeq$ few GeV, for the NFW dark matter density profile. This demonstrates that ORCA- and ARCA-like detectors will be able to perform competitive dark matter searches in a wide range of masses. The implications of these striking projections are investigated in a few selected dark matter particle models, where we show that neutrino telescopes are able to probe new parameter space. 
}

\begin{document}

\maketitle
\flushbottom

\section{Introduction}

\label{sec:intro}

Despite the evidence for existence of large non-baryonic matter content of the Universe~\cite{bertonehooper} the true nature of dark matter (DM) remains a major puzzle in cosmology, astrophysics, and particle physics. 
There are various models predicting DM candidates differing in orders of magnitude in their masses~\cite{AlvesBatista:2021gzc}. Weakly Interacting Massive Particles (WIMPs) are among the most popular candidates and have been extensively tested in different probes of direct and indirect detection and at accelerators. Despite the large efforts being made in last decades, no DM candidate has been identified by any of the search methods, which are particularly sensitive to channels where the DM particles interact with quarks and/or charged leptons.

Among the searches, indirect detection using neutrinos is promising because they travel directly from the source, largely unaffected. This benefit however, requires very sensitive detectors, with detection volumes much larger than $\gamma$-ray telescopes. For this reason the neutrino channel is poorly constrained since generic experiments do not have the sensitivity to detect large amounts of neutrino events. 

The next-generation neutrino experiments and upgrade of currently operational experiments for sub-GeV and light-DM masses will significantly improve on this limitation by featuring large effective volumes. The landscape of these experiments is given by \emph{e.g.} IceCube-Gen2~\cite{Aartsen_2021} and Baikal-GVD~\cite{Baikal-GVD:2019kwy} in the high-DM mass region and Super-Kamiokande~\cite{TAKEUCHI2020161634}, Hyper-Kamiokande~\cite{Hyper-Kamiokande:2018ofw} and DUNE~\cite{DUNE:2020ypp} at low masses. 
Experiments in the northern hemisphere will additionally augment the sensitivity to detect DM signals originating from the Galactic Centre, where DM particles are expected to have the largest density, and will work towards probing the thermal relic scenario for WIMP-like particles.  

Among the next generation experiments we take as benchmark the KM3NeT neutrino telescope~\cite{KM3Net:2016zxf}, which is a new generation cubic kilometer network of deep-sea telescopes located in the abyss of the Mediterranean Sea~\cite{KM3Net:2016zxf}. 
Its two main sites are already partially operating, with the high-energy ARCA and the low-energy ORCA detectors located near the coasts of Italy and France respectively. 
ORCA is optimised for the detection of atmospheric neutrinos with energies above a few GeV. The main goal of the detector is the measurement of atmospheric neutrino oscillations considering all flavours and the neutrino mass ordering~\cite{KM3Net:2016zxf,KM3NeT:2021ozk}.  
However the experimental setup allows for a variety of other studies that test extensions of the SM, such as neutrino non-standard interactions~\cite{KhanChowdhury:2020qqu}, $\tau$ appearance~\cite{Eberl:2017plv}, sterile neutrinos~\cite{KM3NeT:2021uez} and solar dark matter capture~\cite{Lopez-Coto:2020ovv}.
The ARCA energy range is very similar to the one of the ANTARES and IceCube detectors and allows for the study  of high-energy neutrino physics in a broad range. Of interest for us is the study of DM annihilation from the Galactic Centre, see \emph{e.g.}~\cite{ANTARES:2019svn,Aartsen:2020tdl}.
The continuous development of the ORCA and ARCA sites will make KM3NeT one of the largest neutrino telescopes in the near future. Due to its location, that allows for a good field of view of the Galactic Centre, and its improved angular and energy resolutions for track and shower event topologies, KM3NeT will provide a unique probe of DM annihilation signals coming from the heart of the Milky Way.

The main focus of this paper is to explore this possibility thoroughly for a wide range of DM masses, first in a model independent way and then within the context of simple DM models. In a previous study~\cite{BasegmezDuPree:2021fpo}, we reported the sensitivity of KM3NeT using an angular power spectrum (APS) analysis based on the full configuration of the ARCA detector, which has the proper angular resolution for this method, see \emph{e.g.} Refs.~\cite{Aartsen:2014hva,Dekker:2019gpe} for details. This analysis provides conservative upper limits that are very robust against the variation of the shape of the DM density profile towards the Galactic Centre. Here we employ a maximum likelihood method that incorporates the information of angular, energy resolution and event topology, using the design specifications for a fully functioning KM3NeT-like experiment in Ref.~\cite{KM3Net:2016zxf}. In this analysis we extend the DM mass range to span from few GeV up to tens of TeV, by including both the ORCA detector for low-DM masses up to hundreds of GeV and the ARCA detector for high-DM masses up to $10^5$ GeV. We show that the inclusion of full event information improves significantly the sensitivity of the detectors, which display promising prospects for detection of DM annihilation signals. In particular, in case of annihilation into mono-chromatic neutrino lines KM3NeT will be capable of reaching for the first time the thermal relic cross-section for a wide range of DM masses. These prospects nicely complement and/or improve recent studies on neutrino lines sensitivities from DM annihilation, see \emph{e.g.} Refs.~\cite{Arguelles:2019ouk,Bell:2020rkw}. Following the models studied in our previous work~\cite{BasegmezDuPree:2021fpo}, we show the substantial gain in probing the model parameter space via neutrino final states with respect to gamma-ray signals and direct detection experiments. This is particularly true for the gauged  $U(1)_{L_\mu-L_\tau}$ model that is an anomaly free DM models accommodating naturally several anomalies reported in the muon sector and flavor physics and connecting neutrino and DM physics~\cite{PhysRevD.44.2118, PhysRevD.43.R22, Altmannshofer:2016jzy, Bauer:2018onh, Foldenauer:2018zrz, Asai:2020qlp, Heeck:2022znj}.

The rest of the paper is organised as follows. In Sec.~\ref{sec:nusignals}, we remind the reader about the expected neutrino flux from DM and the main background processes. Section~\ref{sec:nutelescope} describes the modelling of the ORCA and ARCA detectors and the statistical analysis that employs a maximum-likelihood ratio method based on mock simulated data-sets. In Sec.~\ref{sec:results} we provide the model-independent upper limits on the DM thermally averaged annihilation cross-section for both the low- and high-energy detection capabilities of the experiment, which covers a large DM mass range. We further discuss the experimental and theoretical uncertainties affecting the upper limits.  We then interpret the forecasts for a KM3NeT-like detector in terms of specific models, as detailed in Sec.~\ref{sec:models}. We conclude with final remarks and possible future studies in this direction in Sec.~\ref{sec:concl}. 

\section{Neutrino flux: signal and background}
\label{sec:nusignals}
\subsection{Dark Matter signals in neutrino telescopes}

DM particles~($\chi$) can annihilate in the galactic halo into Standard Model (SM) particles, which subsequently decay and shower producing neutrinos among the stable final states. The expected neutrino intensity from this annihilation is
\begin{align} \label{eq:dm_flux}
\frac{dI_{\nu}(\psi)}{dE}=\frac{\langle \sigma v \rangle}{8 \pi m_{\rm \chi}^2}\frac{dN_{\nu}}{dE} \, \int_{0}^{l_{max}}\rho^2_{\chi}[r(l)]dl\,,
\end{align}
where $m_{\chi}$ and $\langle \sigma v \rangle$ are respectively the DM mass and the thermally averaged annihilation cross-section, $dN_{\nu}/dE$ is the neutrino energy spectrum per DM annihilation and $\rho_{\rm DM}$ is the DM density distribution.\footnote{Notice that here we consider self-annihilating DM particles; when particle and anti-particle are distinct, as in the case of the DM models discussed in Sec.~\ref{sec:models},  an additional factor of $1/2$ has to be added to Eq.~\eqref{eq:dm_flux}.} The parameters $r$ and $l $ represent the galactocentric and the line-of-sight distances and are related by $r=\sqrt{R_{\oplus}^2+l^2-2Rl\cos\psi}$,
where $\psi$ is the angle between the Galactic Centre and the sky position of interest and $R_{\oplus}=8.5$ kpc is the distance from the center of the galaxy. The line-of-sight integral is often called the J-factor, $\cal J$, in the literature, and it encapsulates the astrophysical dependence of the DM signal in Eq.~\eqref{eq:dm_flux}. The J-factor integral is cut off at the virial radius, R$_{\rm vir}\sim$ 200 kpc, where $l_{max}=R_{\oplus}\cos\psi+\sqrt{R_{\rm vir}^2-R_{\oplus}^2\sin^2\psi}$.

As shown in Eq.~\eqref{eq:dm_flux}, the expected DM flux depends on the density profile squared, and thus the profile choice significantly affects the signal intensity. For our benchmark case, we consider the Navarro-Frenk-White (NFW) profile~\cite{Navarro:1996gj}
\begin{align}
\label{eq:NFW}
\rho_{\chi}(r)=\frac{\rho_{s}}{(r/r_{s})(1+r/r_{s})^2}\,,
\end{align}
where $r_{s}=20$ kpc is the scale radius and $\rho_{s}$ is the scale density, which is fixed by the local DM density $\rho_{\chi}(R_\oplus)$. The local density can be determined by various methods, and its mean value is bracketed between $\left(0.3-0.6\right)$~$\rm GeV cm^{-3}$~\cite{deSalas:2020hbh}. For our results, we set $\rho_{\chi}(R_\oplus) = 0.4~{\rm GeV cm^{-3}}$, and the final results scale with the local density squared for the given profile. We discuss the dependence of our results with respect to the density profile choices in Sec.~\ref{sec:uncertainties}.
Lastly, since we consider annihilating DM, the main region of interest is around the Galactic Centre with a maximum opening angle of $30^\circ$. In this narrow region of interest, the extra-galactic DM flux is negligible and hence not included in the analysis. 

In the case of DM annihilating into two SM particles, the neutrino energy spectra per annihilation required in Eq.~\eqref{eq:dm_flux} are taken from \texttt{PPPC4DMID}~\cite{Cirelli:2010xx} in the low-mass regime ($m_{\rm \chi}< 1$\,TeV) and from \texttt{HDMSpectra}~\cite{Bauer:2020jay} in the high-mass regime ($m_{\chi}> 1$\,TeV). The latter code has an improved handling of the electroweak corrections that could affect significantly the shape of the neutrino energy spectra in the multi-TeV regime, see Ref.~\cite{Ciafaloni:2010ti}. For the annihilation channel directly into neutrinos, $\chi \chi \rightarrow \nu \bar{\nu}$, the energy spectra are simply given by a monochromatic line with analytic expression $dN_{\nu}/dE = 2\delta(m_\chi - E_{\nu})$, where  $\delta$ is the Dirac-delta function. We also consider DM annihilating into 4 leptons ($l$), a final state that can be originated by a $t$-channel annihilation process mediated by an intermediate meta-stable mediator ($Y$), which is typically much lighter than the DM. In case of charged lepton final state, $\chi \chi \rightarrow YY \rightarrow 2 l^{+} 2 l^{-}$, the spectra are taken from Ref.~\cite{Elor:2015bho} as implemented in \texttt{PPPC4DMID}. For $\chi \chi \rightarrow YY \rightarrow 2\nu 2\bar{\nu}$, the energy spectrum has a box shape~\cite{Ibarra:2012dw,Garcia-Cely:2016pse,ElAisati:2017ppn}, with analytical form $dN_{\nu}/dE = 4 H (m_{\chi} - E_{\nu})/m_{\chi}$, where $H$ is the Heaviside step function. 

The neutrinos produced in the galactic halo oscillate on their path to Earth, and the neutrino spectra at detection, is given by: 
\begin{equation}
\label{eqn:detectedfluxes}
\left(\frac{dN}{dE}\right)_{\beta}= \sum_{\alpha}\left(\frac{dN}{dE}\right)_{\alpha}^{src}P_{\alpha \beta}\,,
\end{equation}
where $\left(\frac{dN}{dE}\right)_{\beta}$ is the oscillated neutrino spectra for  $\nu_{\beta}$ neutrino flux coming from the DM annihilation, and $P_{\alpha \beta}$ is the oscillation probability from $\nu_{\alpha}$ at source to $\nu_{\beta}$ at detection. Before reaching Earth, we consider the neutrino oscillations in vacuum. For these, the oscillation length is much smaller than the typical galactic distance, $\sim$\,kpc, hence the neutrinos propagate as incoherent states of mass neutrinos.
The uncertainties in the neutrino mixing parameters only introduce small variations in the final flux ratio, in comparison with other uncertainties discussed in this work, see \emph{e.g.} Ref.~\cite{Bustamante:2015waa}, and will be neglected in the rest of this article.

\subsection{Neutrino Background} 
\label{sec:bkg_model}

The main background for neutrino searches is dominated by cosmic-ray induced energetic atmospheric muons. By choosing only up-going neutrinos in the experiment, one can suppress this background substantially.
The next most important background is cosmic-ray induced atmospheric neutrinos originating from cosmic-ray collisions with the outer atmosphere. These atmospheric neutrinos dominate until around 100\,TeV, where the astrophysical neutrino background starts to increase. The atmospheric neutrino flux is well-studied and well-measured; we use the atmospheric neutrino flux from Ref.~\cite{Honda:2015fha} and extend it to high energies following the parametrisation from Ref.~\cite{Sinegovskaya:2014pia}. For astrophysical neutrinos, we use the parametrisation of the best power-law fit to the neutrino spectra from Ref.~\cite{Stettner:2019tok}.

\section{Detector modelling and statistical analysis}
\label{sec:nutelescope}

When neutrinos enter the detector, a variety of interactions can take place and lead to different types of event topologies.

Charged-current (CC) interactions originating from $\nu_{\mu}$'s produce a final state muon, which can travel long distances and leave distinctive track signatures. CC interactions from $\nu_{e}$ and $\nu_{\tau}$ produce electrons and taus in the final state. Electrons give rise to electromagnetic showers, while taus can produce both electromagnetic and hadronic showers depending on the decay mode. Therefore, in our analysis, we consider showers from $\nu_e$ and $\nu_\tau$ together (see, however, Ref.~\cite{Li:2016kra}). We note that about 17\% of the time taus decay into muons and thus produce track-like signatures. To account for this, we decrease by 17\% the number of shower events from $\nu_{\tau}$-CC reactions. Since the atmospheric $\nu_\mu$ background is dominant for tracks, the additional track signal from $\tau$-decays does not meaningfully improve the signal-to-noise ratio, and are thus neglected. 

Neutral-current (NC) interactions are produced by all flavors, however, the cross section is smaller than that of CC events. NC events also deposit less energy in the detector due to the final state neutrinos, making them in general subdominant as compared to the corresponding CC contributions; we thus also neglect the NC contribution for both the signal and background. 

In general, for a given neutrino intensity, $dI_{\nu}/dE_{\nu}$, the number of expected events in an experiment at the $i^{\rm th}$ energy and the $j^{\rm th}$ angular bins are given by 
\begin{align}\label{eq:event}
n_{ij}= T_{\rm eff}  \int_{i} dE_{\nu} \int_{j} {\rm vis}(\Omega)\, d\Omega \frac{dI_{\nu}}{dE_{\nu}}A_{\rm eff}\langle e^{-\tau(E_{\nu}, \Omega)}\rangle \,,
\end{align}
where $T_{\rm eff}$ is the effective exposure time, $\rm vis(\Omega)$ is the visibility function, $A_{\rm eff}$ is the effective area as a function of the energy, and $\tau(E_{\nu}, \Omega)$ is the optical depth of the neutrinos when they travel through Earth. 
\begin{figure}[tbp]
\centering 
\includegraphics[width=.6\textwidth]{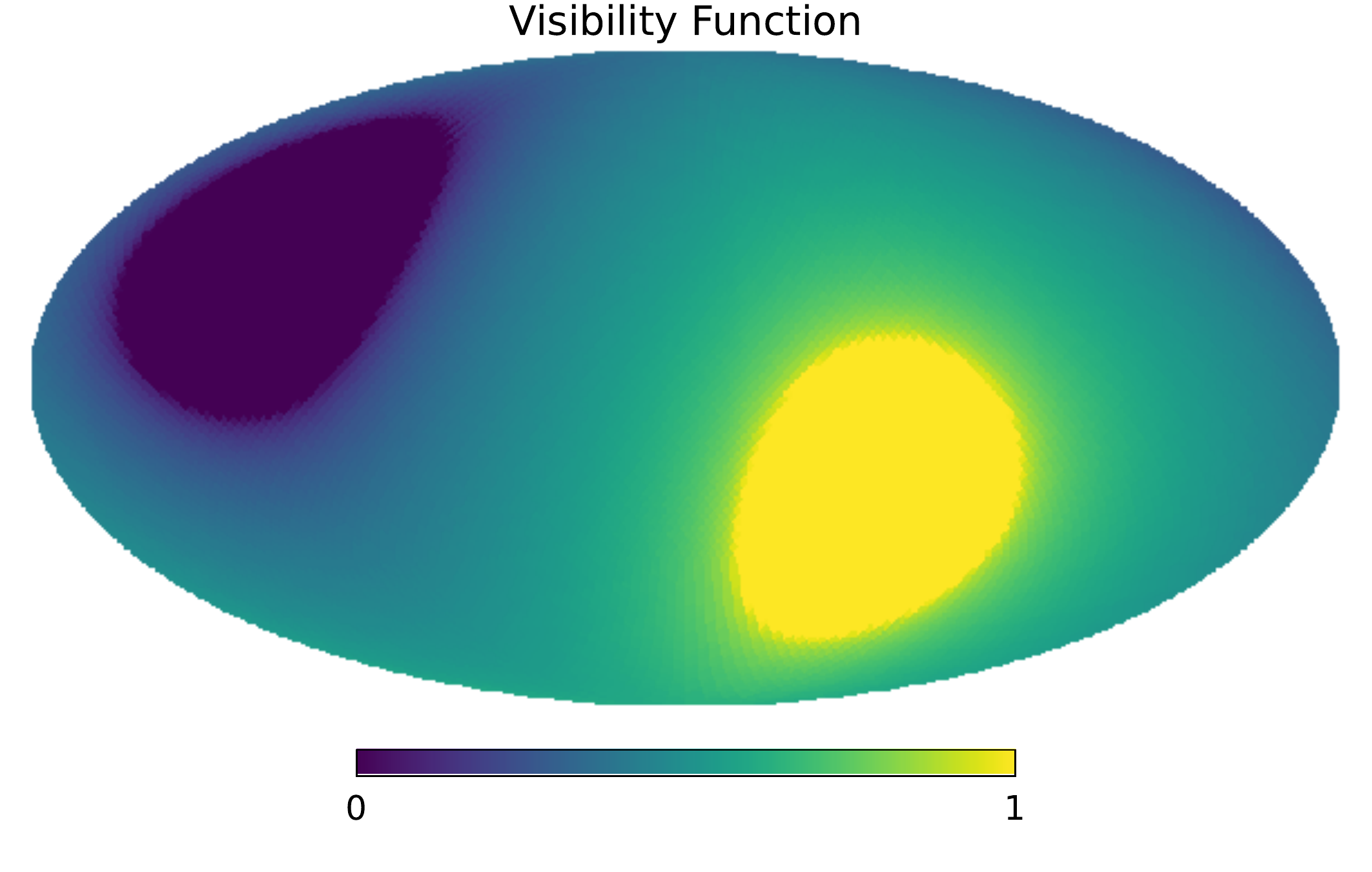}
\includegraphics[width=.6\textwidth]{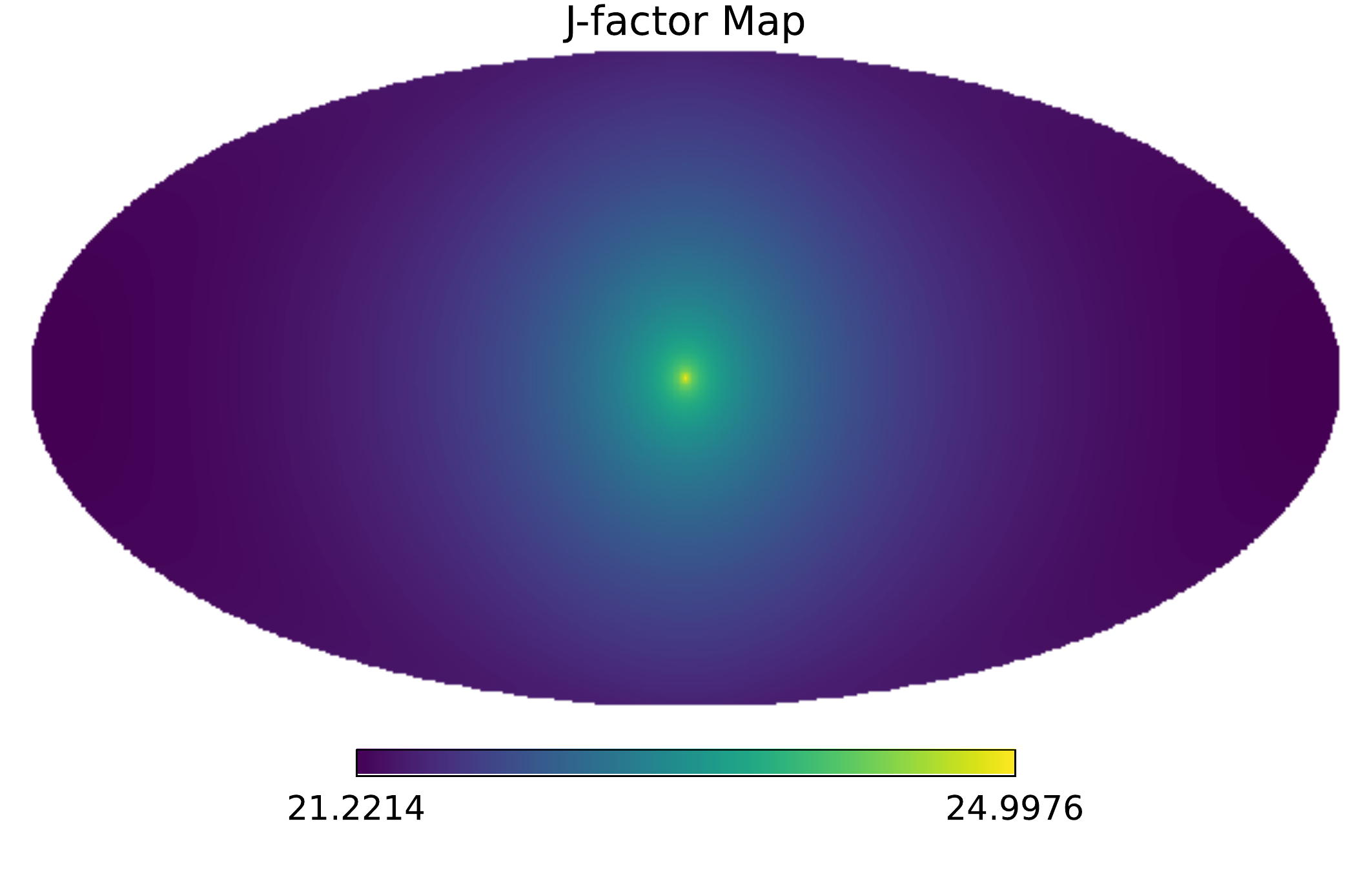}
\includegraphics[width=.6\textwidth]{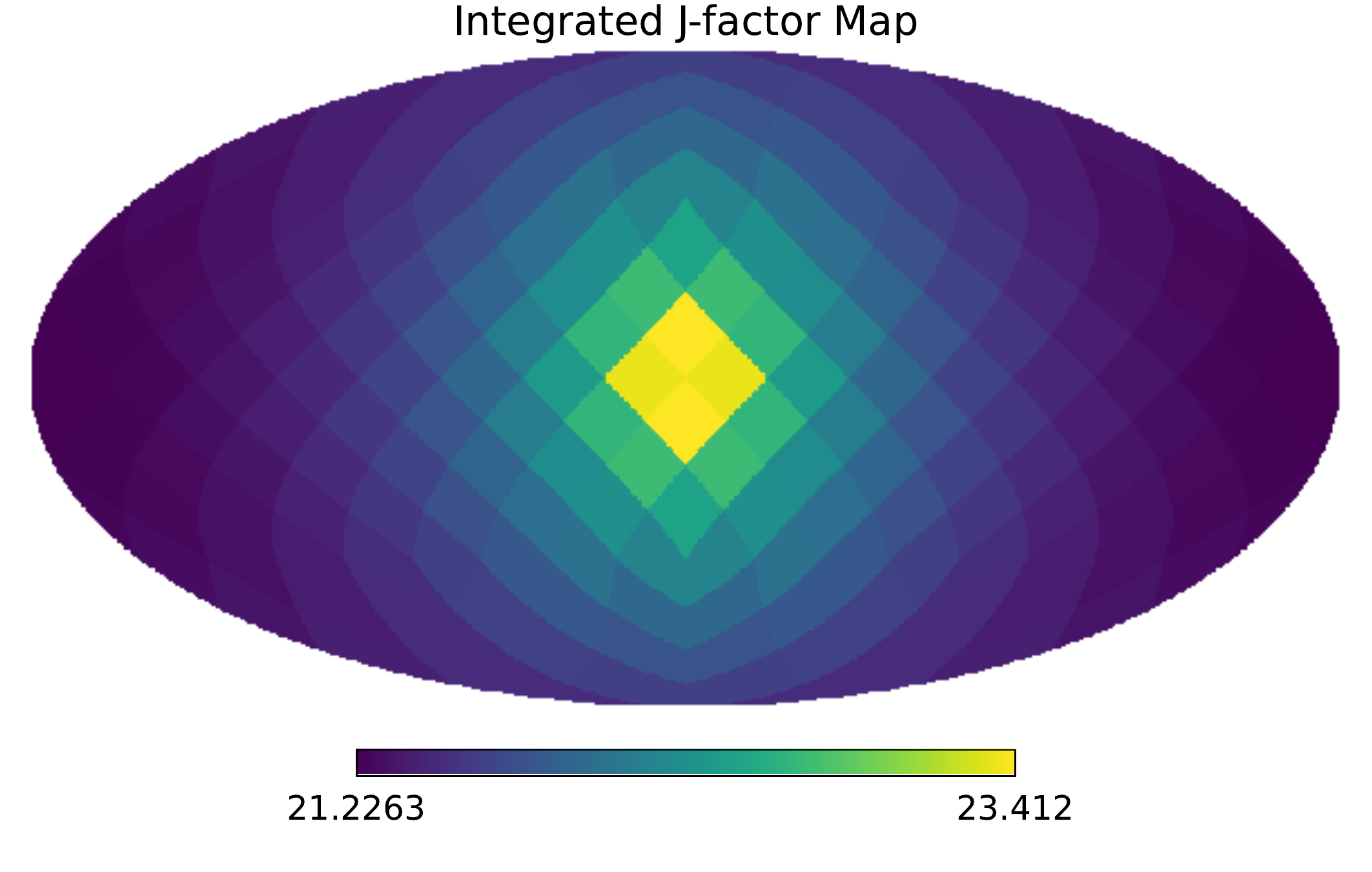}
\caption{Top: Sky map in galactic coordinates convoluted with the visibility function $\rm vis(\Omega)$~ in Eq.~\eqref{eq:vis}, that describes the fraction of time that the sky position is below the detector horizon.  Middle: The J-factor sky map (${\rm log_{10}}{\cal J}$) with a NFW DM density profile, see Eq.~\eqref{eq:NFW}. The DM signal is directly proportional to the J-factor, defined in Eq.~\eqref{eq:dm_flux}. Bottom: The angular averaged J-factor sky map, $\langle {\cal J} \rangle$, where each pixel corresponds to a sky area of about $(200^\circ)^2$. This corresponds to angular resolution of about $7^\circ$ (before multiplying by the factor of 2, as explained in the text.)}
\label{fig:skymaps}
\end{figure}

For our detection setup, we take  $T_{\rm eff}= 10 \,{\rm yrs}$.  We only consider up-going neutrino events to avoid the atmospheric muon background contamination. 
The visibility function  is defined as the fraction of time when the sky position is below the horizon to filter out the atmospheric muon background and encapsulates how different points in the sky are accessible by the detector for differing periods of time. It is necessary to follow the sky positions relative to the detector zenith. We consider the detector at a latitude about $36^\circ$. north, which is roughly the latitude of the KM3NeT experiment. Then for each position in the sky, we track its declination relative to the detector using \texttt{Astropy}~\cite{astropy:2013, astropy:2018} as
\begin{equation}\label{eq:vis}
    {\rm vis(\Omega)} = \frac{1}{T} \sum_{i}{ \Theta(-\delta_{i}(\Omega)) \Delta T } \, ,
\end{equation}
where $\Theta$ is the theta function, $\delta$ is the declination of the sky position $\Omega$ relative to the detector zenith and $T$ = 7 days, which is a time period long enough to average the effect of Earth rotation. 
The energy dependence of up-going neutrinos being absorbed by the Earth is described by the optical depth, $\tau(E_{\nu},\Omega) = n \sigma L$, where $n$, $\sigma$, and $L$ are the nucleon number density, neutrino-nucleon cross-section and neutrino path length respectively~\cite{Gandhi:1998ri}. 
Additionally, for each position in the sky, the neutrino path length varies as the Earth rotates, thus the neutrino absorption factor is averaged with respect to the probability of each path length as $\langle e^{-\tau(E_{\nu}, \Omega)}\rangle$.

Figure~\ref{fig:skymaps} (top panel) shows the visibility function in galactic coordinates: the color code goes from 0 to 1, where the latter stands for a sky position always below the horizon hence always visible to KM3NeT, while the former is considered not visible given the location and the choice of the arrival neutrinos.

We note that the Galactic Centre is visible with a high-time fraction of about 63\% and DM annihilation signals are expected to peak along that direction.  Such a high visibility towards the Galactic Centre is a key advantage of KM3NeT for DM searches over neutrino detectors in the southern hemisphere. 

Figure~\ref{fig:skymaps} (middle panel) shows the J-factor in Eq.~\eqref{eq:dm_flux}, which is directly proportional to the expected DM intensity.  Neutrino telescopes are not fully sensitive in such fine angular distribution to such a J-factor map, as they are limited by their angular resolution.  In order to compute the J-factor integration, we split the map according to the energy and angular resolutions (setups detailed in the next section). Briefly, we first define the energy bins according to energy resolution. Then for each energy bin, a sky map is partitioned according the median angular resolution $\delta \theta$, with pixel sizes roughly given by $\Delta\Omega_{j} = 2 \pi (1-\cos(\delta \theta))$, using \texttt{Healpix}~\cite{healpix:2005ApJ...622..759G, healpy:Zonca2019}. Notice that for both the energy bin widths and the pixel angular size, we take twice of the energy and angular resolution to be conservative. The numerical integration in Eq.~\eqref{eq:event} is then performed by further subdividing each pixel into smaller pixels with a maximum size about $4\times 10^{-6}$\,sr or $(0.01^{\circ})^2$. 

Figure~\ref{fig:skymaps} (bottom panel) shows the angular averaged J-factor $\int d\Omega {\cal J}/\int d\Omega$, for a pixel size of about $(200^\circ)^2$. As can be seen from the map, after integrating over each pixel (as required in Eq.~\eqref{eq:event}), the angular distribution of the signal strength is broadened, and becomes less peaked. This shows how the effect of telescope's angular resolution degrades the signal-to-noise ratio. We note that in this example, the large pixel size is more representative for the case of ORCA, given the superior angular resolution of ARCA a finer pixelisation is used.

\subsection{ORCA}
\label{sec:orca}

ORCA is the part of the KM3NeT neutrino telescope located at 40 kilometers offshore from Toulon in France at a depth of 2450 m. It is currently being constructed and already operational with 11 strings. The final configuration foresees 115 strings with an average spacing between them of 20 m. Each string containing 18 light detection units spaced 9 m apart in vertical distance, culminating in a total volume of $\sim 3.7$ Mton. Once completed, ORCA will be one of the largest neutrino telescopes in the Northern hemisphere sensitive to neutrinos in the GeV energy range. 

All neutrino flavors contribute to signal events in the ORCA detector as tracks or showers. We collect the effective area, energy and angular resolution of the ORCA detector from Refs.~\cite{ KM3NeT:2021ozk, KhanChowdhury:2021kce} and extrapolate to a maximum neutrino energy of 100 GeV. So for the neutrino line signal ($\chi \chi \to \nu \bar{\nu}$) 100 GeV represents the upper DM mass to which our analysis is limited to, while for a continuum neutrino energy spectrum we can extend the sensitivity of ORCA up to $m_\chi = 10^3$ GeV. 

To compute the sensitivity of ORCA for DM annihilation signals, we consider atmospheric neutrinos as the dominant background.  In this energy range, however, neutrino oscillations in matter are important and are included in our analysis to have an accurate estimate of the background. For definiteness we take the neutrino parameters from Ref.~\cite{ParticleDataGroup:2020ssz}, assume $\delta_{\rm CP}=0$ and normal ordering for the neutrino mass hierarchy.

To calculate the transition probabilities neutrino propagating inside the Earth, we follow the formalism presented in Refs.~\cite{Akhmedov_2008,Akhmedov:2012ah}. For our parameter choice, it entails solving numerically the Schrödinger equation with the Hamiltonian:
\begin{align}
H_{f}=\frac{1}{2E}U_{f}MU_{f}^T+V_{f} \,.   
\end{align}
The matter potential matrix in the flavor basis, $V_{f}=\operatorname{diag}(V_{e},0,0)$, takes the form of
\begin{align}
V_{e}=\sqrt{2}G_{F}N_{e}\approx 3.78\times 10^{-14} \left(\frac{\rho}{\rm g/cm^3}\right) \rm{eV}\,,
\end{align}
where $G_F$ is the Fermi constant, $N_e$ is the electron number density in the medium and $\rho$ is matter density calculated from the density profile of the Earth according to
the Preliminary Reference Earth Model in Ref.~\cite{Dziewonski:1981xy}. Notice that for antineutrinos $V_{e} \longrightarrow -V_{e}$.
\begin{figure}[t]
\centering 
\includegraphics[width=.45\textwidth]{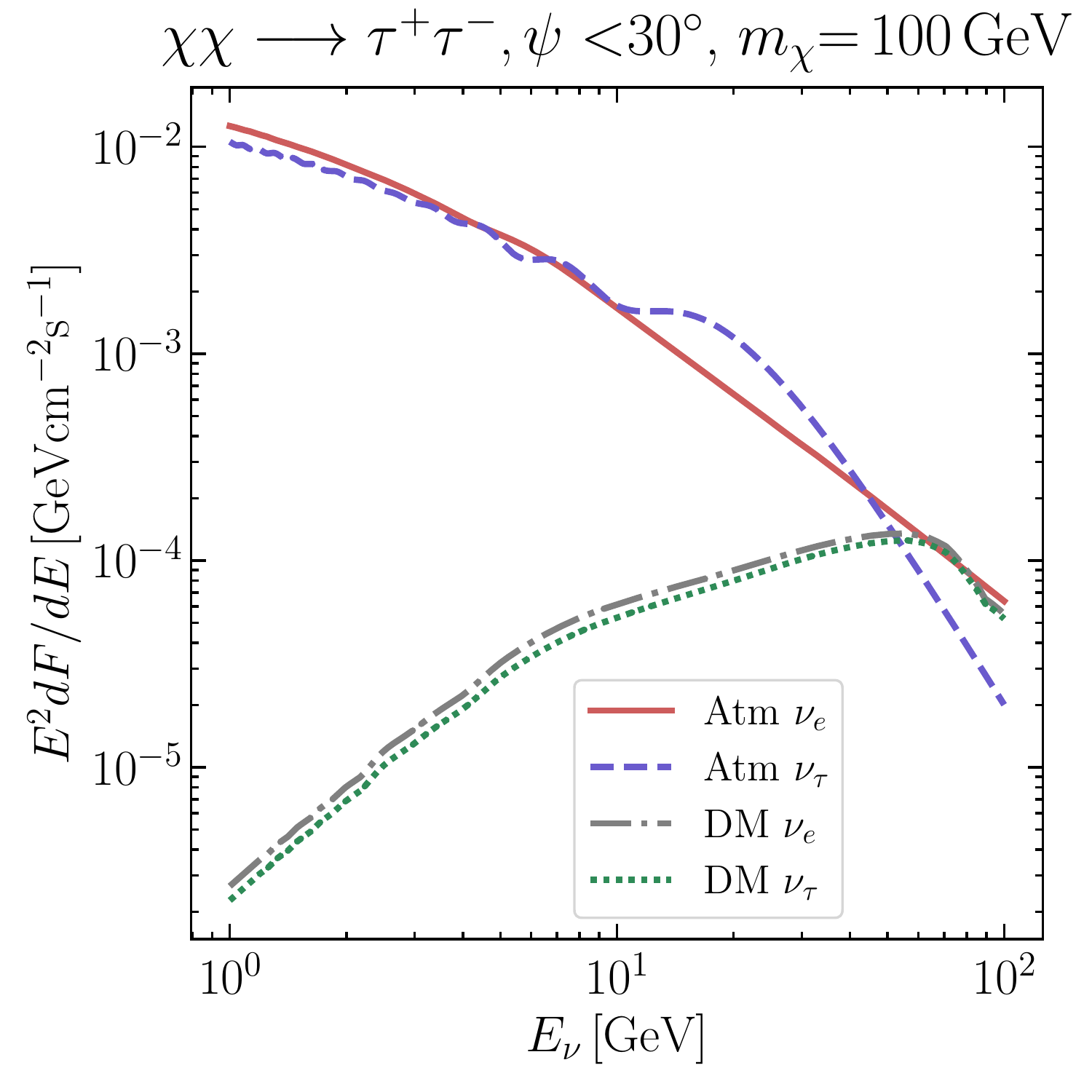}
\hfill
\includegraphics[width=.45\textwidth,origin=c]{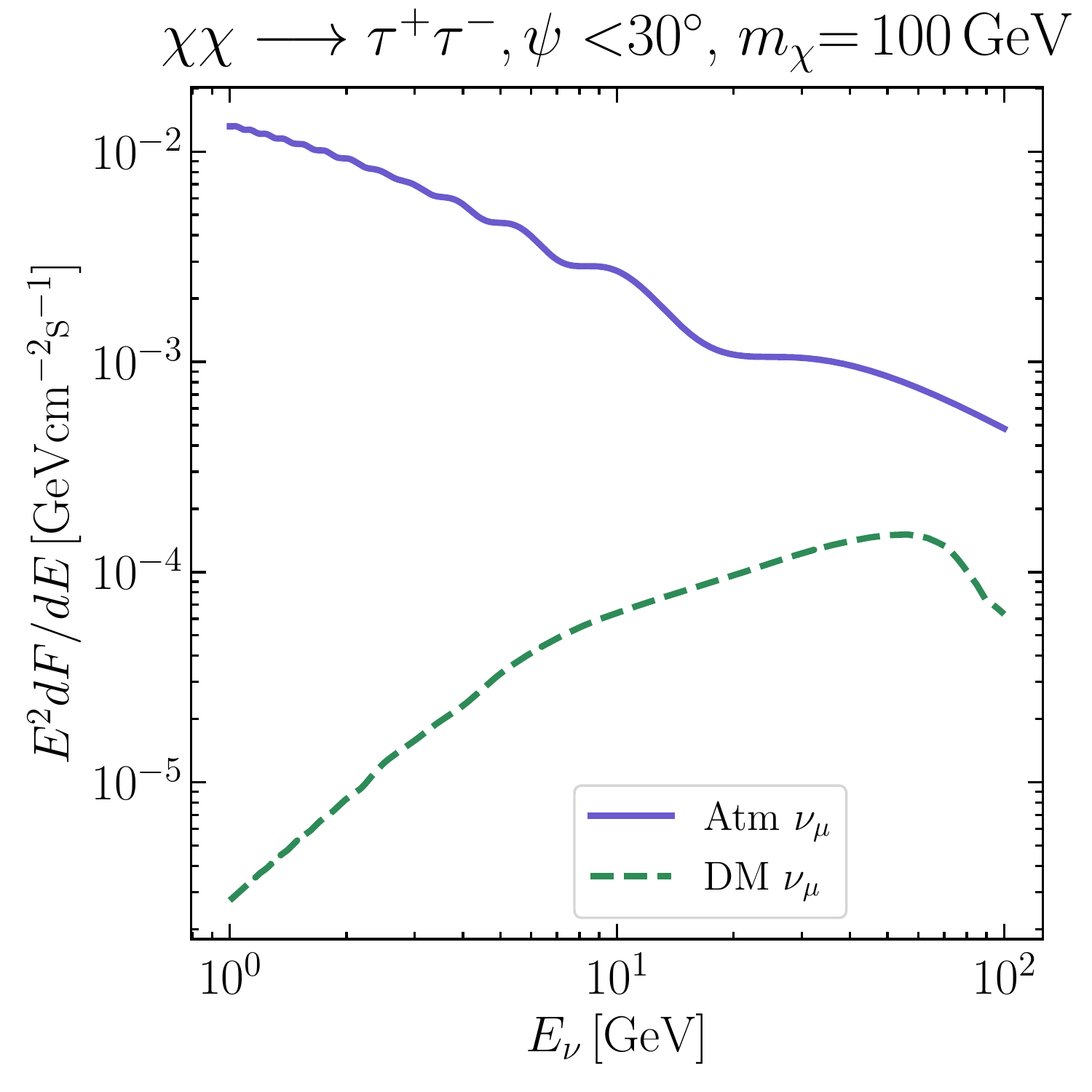}
\caption{Left: Expected $\nu_e$ and $\nu_\tau$ flux (intensities integrated over the solid angle) in the reach of KM3NeT-ORCA including neutrino oscillations in matter on the atmospheric neutrino background, for a DM benchmark given by $\chi \chi \to \tau^+\tau^-$, $\langle \sigma v \rangle=10^{-23}\textrm{cm}^3 \textrm{s}^{-1}$ and $m_\chi = 100$ GeV. The solid red and dotted blue lines correspond to 
($\nu + \bar{\nu}$) electron and tau atmospheric neutrinos, respectively; the dot-dashed and dotted green lines stand for the DM $\nu_e$ and $\nu_\tau$ signals respectively. Right: Same as left for muon neutrinos. The solid blue line denotes the atmospheric background while the green dotted line is for the DM signal, as labelled.} 
\label{fig:fluxes_orca}
\end{figure}
The expected neutrino fluxes measured by the detector are given by Eq.~\eqref{eqn:detectedfluxes}. For the atmospheric neutrinos $P_{\alpha \beta}=P_{\alpha \beta}^{\bigoplus}$ where $\bigoplus$ refers to Earth and for DM neutrinos $P_{\alpha \beta}=\sum_{i} P_{\alpha i}^{src} P_{i \beta}^{\bigoplus}=\sum_{i}|U_{\alpha i}|^2|\sum_{\eta} A_{\beta\eta}^{\bigoplus}U_{\eta i}|^2$~\cite{Razzaque:2010kp}, including the incoherent propagation of neutrinos in space plus the Earth propagation effects. 
We further average the neutrino probabilities in Eq.\eqref{eqn:detectedfluxes} with respect to the azimuthal angle between Earth axis and detector location as $\cos\theta_{z} \in$ [-1,0] therefore we use the angle averaged  $\langle P_{\alpha \beta} \rangle_{\theta_{z}}$ which provides an average flux from upgoing neutrinos. 

Figure~\ref{fig:fluxes_orca} depicts the expected neutrino flux from DM annihilation as well as from the atmospheric neutrino background, for an opening angle of $\psi=30^\circ$  including neutrino oscillations. It can be seen that neutrinos produced by DM are averaged and basically equally mixed when they arrive at the detector, due to their propagation through the intergalactic medium in the galaxy. This is of course not specific to the benchmark chosen ($\chi \chi \longrightarrow \tau^+ \tau^-$ channel with $\langle \sigma v \rangle=10^{-23}\textrm{cm}^3 \textrm{s}^{-1}$ and $m_\chi =100$ GeV) but it is due to large distances from the Galactic Centre. On the other hand, atmospheric neutrinos are sensitive to oscillations in matter, as expected. The $\nu_\tau$ and $\nu_{\mu}$ neutrinos are equally mixed up to energies of $E_\nu \simeq 50$ GeV, while $\nu_e$ are affected for energies $E_\nu \lesssim 20$ GeV and manifest a change in slope in their spectrum. We expect that neutrino oscillations affect muon neutrino flux for energies with a maximum $\sim 50$ GeV. For higher energies, matter effects from neutrino oscillations through Earth are negligible. 

\subsection{ARCA}
\label{sec:arca}

The ARCA site of the KM3NeT neutrino telescope is located about 3500 m at the deap-sea and 100 km offshore from Porto Palo di Capo Passero in Sicily, Italy. It consists of the same optical modules of ORCA but coarsely configured in a vertical distance of 36 m  and average line space of 90 m totaling a half cubic kilometer of instrumented volume. In our studies we consider two identical 115 line configurations so-called the two building blocks of the experiment. ARCA is dedicated to high-energy cosmic neutrino detection with a large energy range starting from 100 GeV, see Ref.~\cite{KM3NeT:2018wnd}.  

Due to its specifications, ARCA is able to detect $\nu_{\mu}$ up to energies of $10^6$\,GeV and we choose to study its sensitivity to DM annihilation in the mass range from 200 GeV up to 10$^5$ GeV, value dictated by the unitarity bound for the thermally averaged cross-section on a $s$-wave annihilating DM particle~\cite{Griest:1989wd, Smirnov:2019ngs}. 

As for the neutrino signal, we consider only track event topologies, as the event and angular resolution information are not available for cascades from Ref.~\cite{Adrian-Martinez:2015wey}. We leave the inclusion of cascades in the ARCA energy range for future works, see {\emph e.g.} Ref.~\cite{ANTARES:2019svn}. 
We consider the effective area and the angular resolution from Figs.~19 and~22 respectively of Ref.~\cite{KM3Net:2016zxf}. 
We note that in the low energy range the angular resolution is dominated by the $\nu_{\mu}-\mu$ opening angle. We thus extend this plot from 0.5 to 0.2 TeV following an $1/\sqrt{E_{\nu}}$ extrapolation. For the energy resolutions, we adopt the value of $\Delta \log_{10}(E_{\nu}) = 0.27$. We discuss in Sec.~\ref{sec:uncertainties} how our results would change under different variations of these specifications. 

Atmospheric neutrinos are the dominant background for DM searches. Although astrophysical neutrinos becomes more dominant above 100 TeV, they however are the same order of the background around 100 TeV and are equally considered in the analysis. We thus consider both atmospheric and astrophysical neutrino backgrounds in all energy ranges as discussed in Sec.~\ref{sec:bkg_model}. In the high-energy range of ARCA oscillation in matter of the atmospheric neutrinos can be ignored, as explained in~\ref{sec:orca}, however neutrino absorption trough Earth becomes important and is considered by the optical-depth factor, as explained below Eq.~\eqref{eq:event}.

\subsection{Maximum-likelihood ratio analysis method}
\label{sec:results_model_indep}

We can compute the expected number of events for both the DM signals and the background in each angular and energy bin by means of Eq.~\eqref{eq:event}. This allows us to perform a binned maximum-likelihood analysis to estimate the expected sensitivities of KM3NeT in terms of upper limits on the thermally-averaged DM annihilation cross-section, $\langle \sigma v\rangle$.
\begin{figure}[t]
\centering 
\includegraphics[width=.45\textwidth]{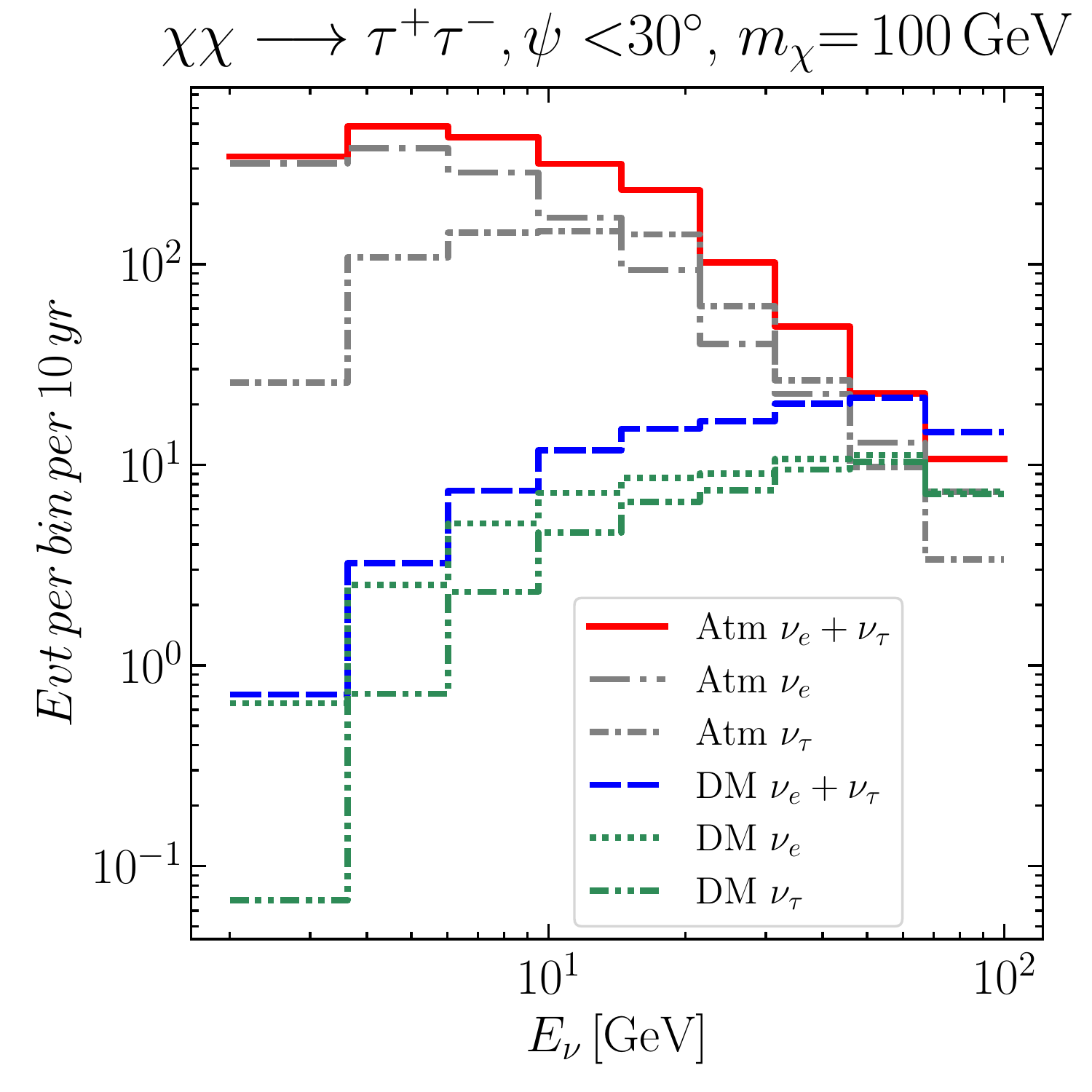}
\hfill
\includegraphics[width=.45\textwidth,origin=c]{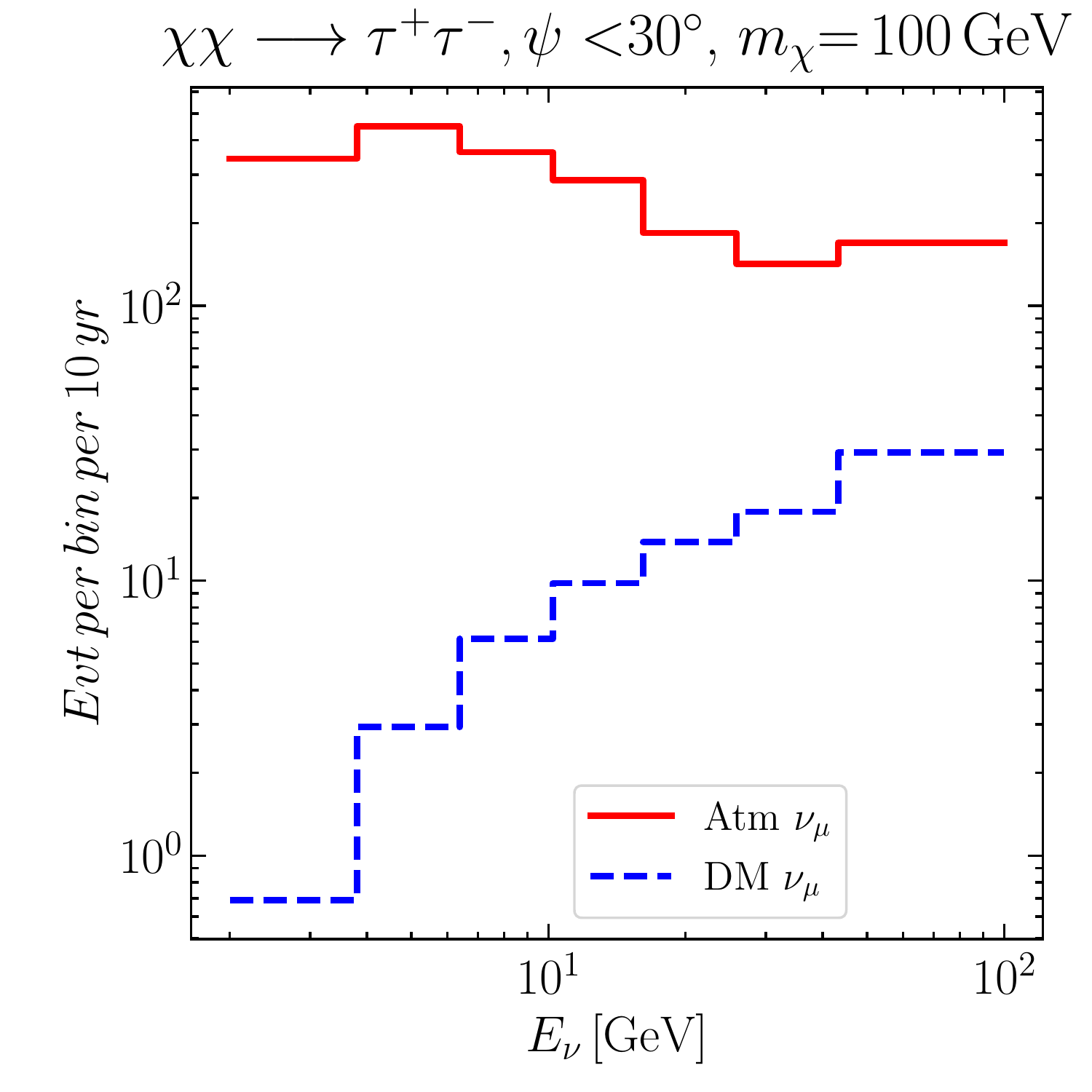}
\includegraphics[width=.45\textwidth,origin=c]{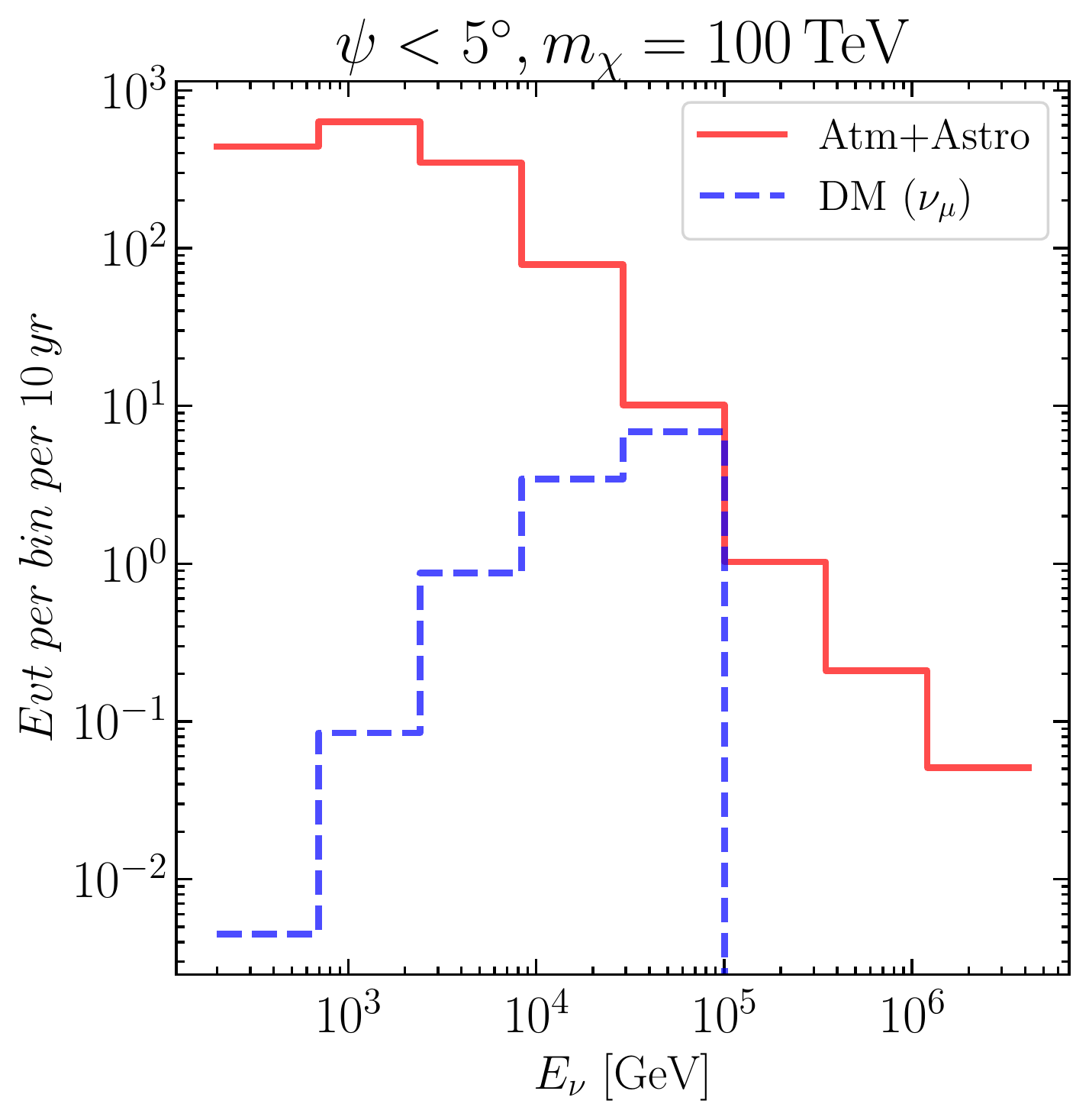}
\caption{Upper left: Number of expected events from electron and tau neutrinos in the ORCA detector for the DM annihilation channel of $\chi \chi \longrightarrow \tau^+ \tau^- $, with an opening angle $\psi=30^\circ$ for $m_\chi=100$ GeV and $\langle \sigma v\rangle=10^{-23}\textrm{cm}^3 \textrm{s}^{-1}$. The dashed blue line shows the contribution from DM neutrinos for both flavors combined, while the solid red line shows the number of events coming from the atmospheric neutrinos ($e$ and $\tau$ flavours summed). Upper right: Same as the left for muon neutrinos. The dashed blue line depicts the events originated by the DM annihilation, while the red solid line shows the events expected from the atmospheric muon neutrino background. 
Bottom: Same as upper right panel for the ARCA detector with an opening angle of $\psi <5^\circ$, with DM parameters $m_\chi=100$ TeV and $ \langle \sigma v \rangle=10^{-24}\textrm{cm}^3 \textrm{s}^{-1} $. }
\label{fig:number_events}
\end{figure}

The likelihood function is given by the Poisson distribution:
\begin{align}
\textrm{L}( \langle \sigma v\rangle )=\prod_{ij} \left[ \frac{\mu_{ij}^{n_{ij}} e^{-\mu_{ij}}}{n_{ij}!} \right] \, ,
\end{align}
where $n_{ij}$ is the number of events in a given energy bin of $i$ and an angular bin of $j$. The $\mu_{ij}(\langle \sigma v\rangle) = S_{ij}(\langle \sigma v\rangle) + B_{ij}$ is the expected number of events for the  DM signal~($S_{ij}(\langle \sigma v\rangle)$) and background~($B_{ij}$) components. For each DM mass hypothesis, the parameter of interest $\langle \sigma v\rangle$, is directly proportional to the DM signal strength. 

The expected number of background and DM annihilation signal events, assuming a 10 year run time is shown in Fig.~\ref{fig:number_events} for ORCA and ARCA separately. The DM signal is produced by $\chi \chi \longrightarrow \tau^+ \tau^-$ channel with $\langle \sigma v \rangle=10^{-23}\,\textrm{cm}^3\,\textrm{s}^{-1}$, $m_\chi =100$ GeV and for a maximum opening angle of $\psi=30^\circ$\ for ORCA, and $\langle \sigma v \rangle=10^{-24}\,\textrm{cm}^3 \,\textrm{s}^{-1}$, $m_\chi =100$ TeV, a maximum opening angle of $\psi=5^\circ$ using tracks for ARCA. For ORCA, the signal-to-background ratio for the shower channels ($\nu_{e}$ and $\nu_{\tau}$) appears to be higher compared to tracks ($\nu_{\mu}$) due to the lower background rate. In the full analysis, the track channel would benefits from better angular resolutions when considering angular bins from the Galactic Centre. In case of ARCA, we note that both atmospheric and astrophysical neutrino backgrounds are included.

For each annihilation channel, we compute the test statistics (TS) to obtain the expected sensitivity, following the description in Ref.~\cite{Cowan_2011}, as
\begin{align}
\textrm{TS}(\langle \sigma v \rangle)=-2\,\textrm{In}\left[\frac{\textrm{L}(\langle \sigma v \rangle)}{\textrm{L}(0)}\right] \,.
\end{align}
 
According to Wilk's theorem~\cite{Wilks:1938dza}, the test statistics under the null hypothesis is $\chi ^{2}$ distributed. In this work we consider the sensitivity at one-sided 95$\%$ Confidence Level (CL) that corresponds to TS$(\langle \sigma v \rangle)=2.71$. We obtain the estimated upper-limit for individual DM annihilation channels with ORCA and ARCA detectors separately. For ORCA, we combine statistically the shower and track channels by considering a joint likelihood, \emph{i.e.}, ${\rm L}_{\rm total} = {\rm L}_{\rm shower}{\rm L}_{\rm track} $.  For ARCA, as discussed previously, only tracks are considered. 
To obtain the statistical errors due to variations, we obtain $n_{ij}$ by Monte Carlo (MC) sampling with the background only hypothesis (\emph{i.e.}, $B_{ij}$) with Poisson probability distribution function. The median expected limits, as well as 68\% and 95\% range of the limits are obtained from 500 MC data sets.

\section{Results and discussion}
\label{sec:results}
\begin{figure}[t]
\centering 
\includegraphics[width=.45\textwidth]{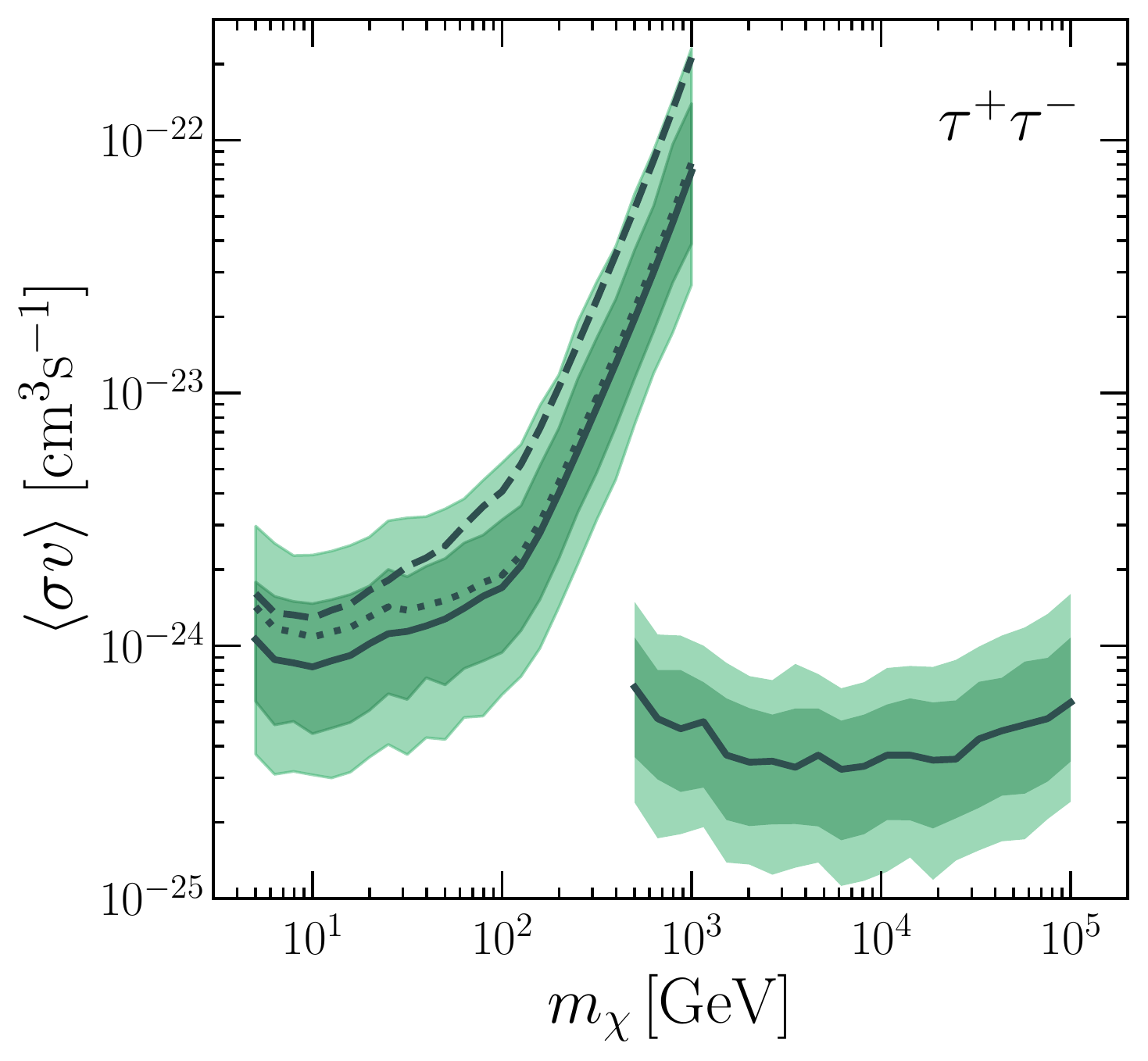}
\hfill
\includegraphics[width=.45\textwidth,origin=c]{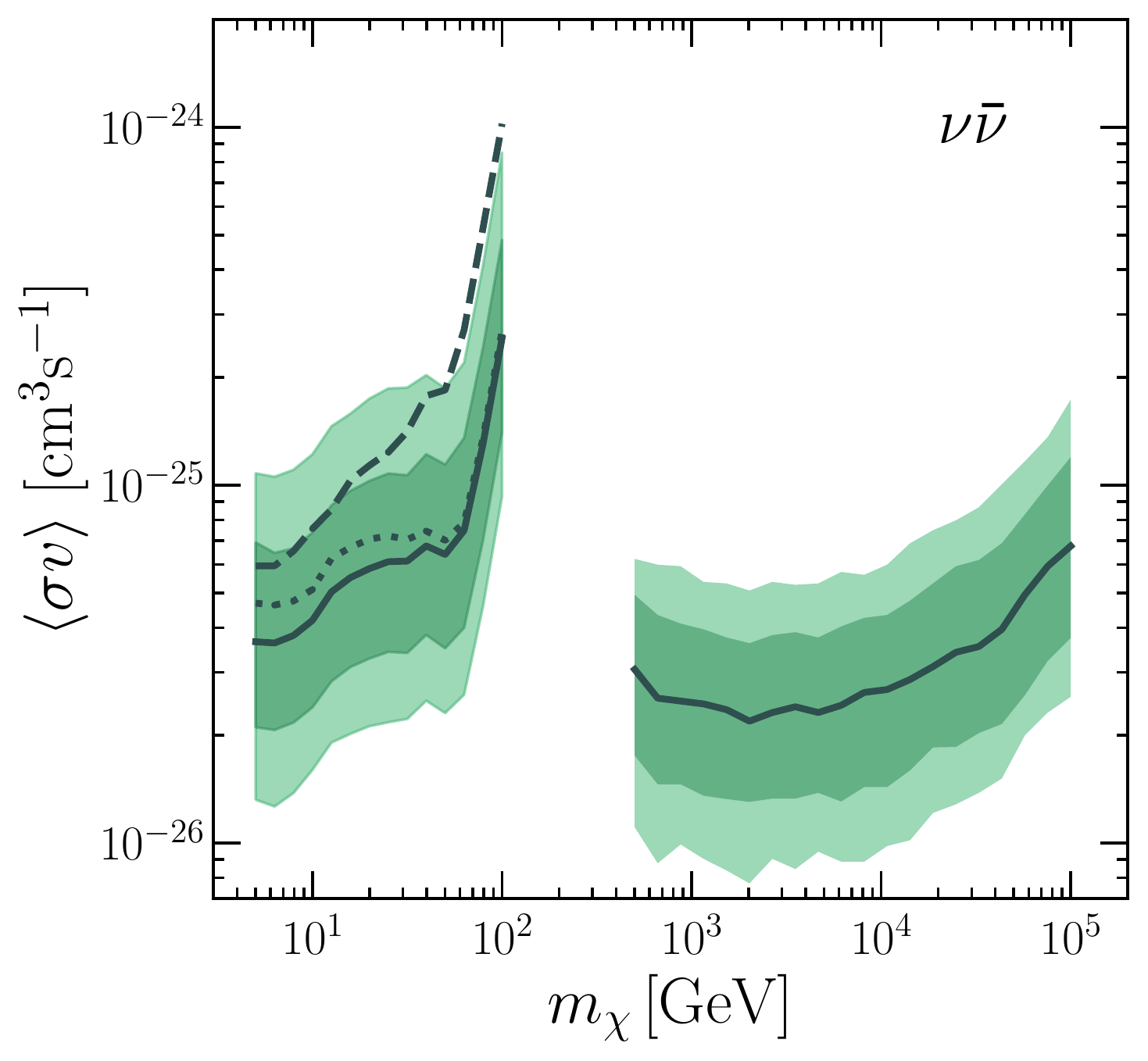}
\caption{Left: Sensitivity of ORCA and ARCA for the DM annihilation channel into $\tau^+ \tau^-$. For ORCA both tracks (black dashed) and showers (black dotted) are shown separately and together as total expected sensitivity (black solid). In the case of ARCA only track events are used, given by the solid black line. The green shaded bands represent the 68\% (dark) and 95\% (light) intervals around the expected value. Right: Same as left for the $\nu \bar{\nu}$ annihilation channel.} 
\label{fig:main_sensitivities}
\end{figure}

We present the sensitivity of the ORCA and ARCA detectors to DM particles annihilating into some specific SM final states, which are assumed to have a 100\% branching ratio one at a time. More precisely we consider annihilation into leptons (2 and 4 charged $l^\pm$) as they feature hard energy neutrino spectra, on the contrary to quarks which have softer spectra. For the latter we only discuss the $b\bar{b}$ final state as representative of coloured fermions. We consider then spectral features such as lines ($\nu\bar{\nu}$) and boxes ($4 \nu$). The upper-limits for gauge boson final states, such as  $W^+ W^-$, are even less sensitive that the case of quarks and are not shown here since we do not use it in the model interpretation.

\begin{figure}[t]
\centering 
\includegraphics[width=.45\textwidth]{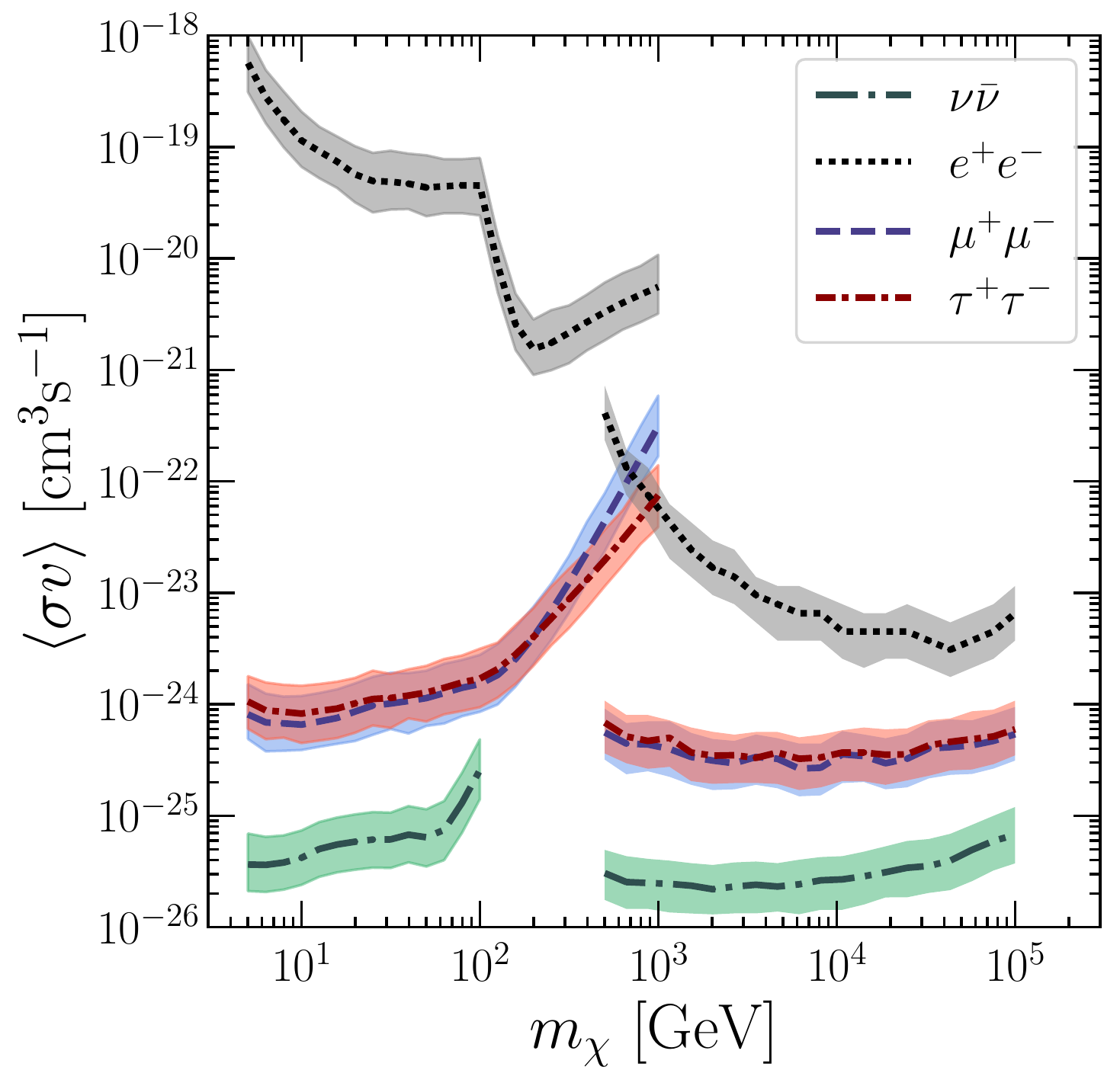}
\hfill
\includegraphics[width=.45\textwidth,origin=c]{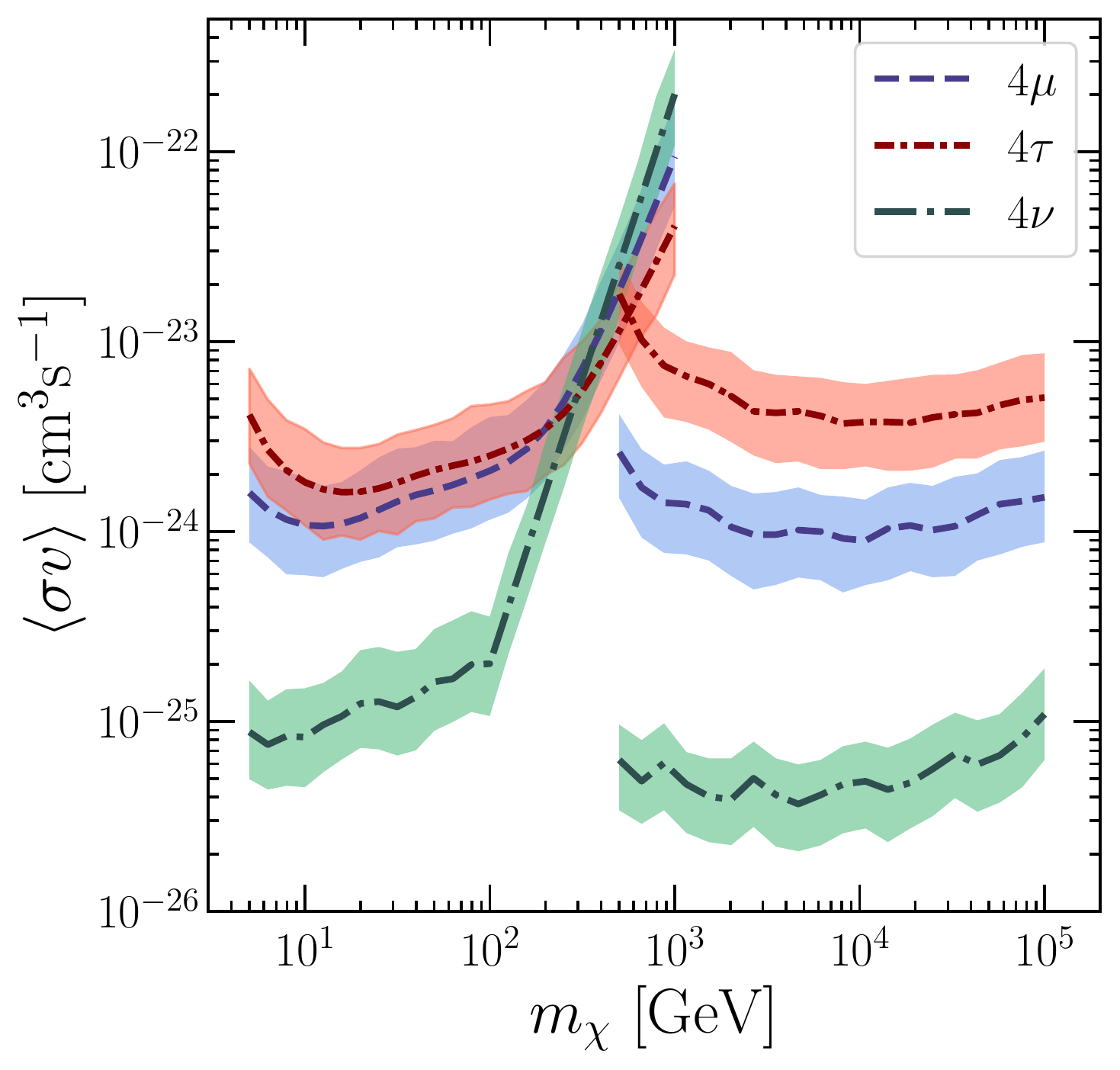}
\caption{Left: ORCA and ARCA expected sensitivity for DM annihilation into two-lepton final states, as labelled. The light colored bands denote the 68\% interval around the expected value, given by the long dot-dashed, dashed, dot-dashed and dotted line for $\nu \bar{\nu}$, $\mu^+\mu^-$, $\tau^+\tau^-$ and $e^+e^-$ respectively. Right: Same as left for the four-lepton final states, as labelled. }
\label{fig:all_sensitivities}
\end{figure}
Figure~\ref{fig:main_sensitivities} shows the expected upper limits for the $\tau^+\tau^-$ and $\nu \bar{\nu}$ channels obtained with the maximum-likelihood analysis, for the low- and the high-mass DM mass range accessible to ORCA and ARCA. The black lines are the expected values while the dark and light green bands denote the 68\% and 95\% intervals, taking into account the statistical uncertainties via MC simulations. For ORCA, the dotted (dashed) line is obtained from shower (track) events only, while the solid line is the sensitivity combining both event topologies. Clearly shower events offer the best sensitivity for ORCA and drive the total upper limit. In case of ARCA, we have to rely on track events only, as explained previously. Notice that in the case of $\nu \bar{\nu}$, KM3NeT is capable of reaching the freeze-out value for the thermally averaged annihilation cross-section. There is a gap in the sensitivity between 100 GeV and 500 GeV due to the different energy coverage of ORCA and ARCA. The hard spectral shape of the $\nu$ and $\tau$ channels also explains the rapid lose in sensitivity registered in ORCA once the DM mass reach beyond 100 GeV in the energy range. As can be seen, there is a slight worsening in the ORCA sensitivities around the mass of 20 GeV for $\nu \bar{\nu}$ channel. This effect arises from neutrino oscillations in matter that increase the number of expected background events. Indeed, as shown in Fig.~\ref{fig:fluxes_orca} there is an interplay between the muon and tau flavour oscillation resulting to a slight increase of the atmospheric $\nu_\tau$ contribution. Therefore, it is important to carefully include the effect of neutrino oscillation in the low-mass DM searches.
\begin{figure}[t]
\centering 
\includegraphics[width=.9\textwidth]{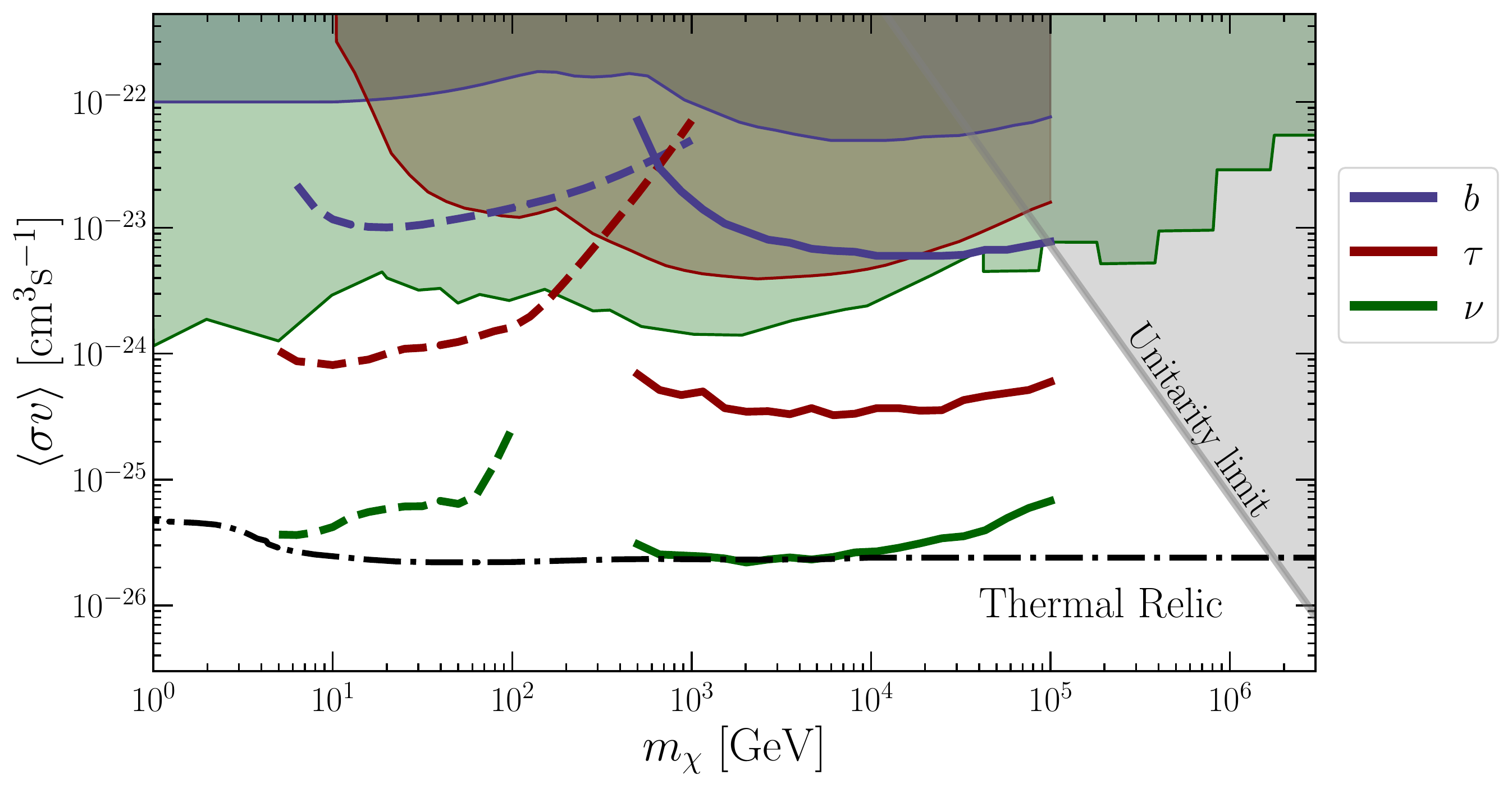}
\caption{The expected sensitivity reach of KM3NeT ORCA (dashed lines) and ARCA (solid lines) for $b\bar{b}$, $\tau^{+}\tau^{-}$ and $\nu\bar{\nu}$ annihilation channels, which are representative cases for soft-hadron, hard-leptonic and the neutrino-line channels. For comparison, we also show the current upper limits from different analyses taken from~\cite{Arguelles:2019ouk}. The shaded green region is the exclusion bound for the $\nu \bar{\nu}$ channel, while the red and blue shaded regions are up-to-date exclusion bounds for annihilation into $\tau^+ \tau^-$ and $b\bar{b}$ respectively. We find that, in general, KM3NeT will offer substantial improvements to the DM sensitivity compared to existing results. In particular, the neutrino-line channel may potentially reach the thermal relic~\cite{Steigman:2012nb} level (dot-dashed black line), which is an important milestone for testing the WIMP DM hypothesis. The shaded grey region in the right top corner violates unitarity of the annihilation cross-section. }
\label{fig:main_plot}
\end{figure}

Figure~\ref{fig:all_sensitivities} shows the expected upper limits for all leptonic channels considered in the analysis, as labelled, with only 68\% statistical error bands. In the left panel, we note that the $e^+ e^-$ channel has a sensitivity boost at around $100\,{\rm GeV}$, which is caused by the neutrino spectral hardening from the PPPC4DMID~\cite{Cirelli:2010xx} spectrum data bank.
The $\mu^+\mu^-$ and $\tau^+ \tau^-$ channels perform similarly, while the $\nu\bar{\nu}$ channel has
the best sensitivity, as expected.
Considering the case of four leptons (right panel), we do not show the $4 e$ case as the sensitivity is too weak. We remark a distinct behavior between the $4 \tau$ and $4 \mu$ channel, the latter begin a more optimal signal for the ARCA detector. The $2$- and $4$-lepton energy spectra are taken respectively from \texttt{HDMSpectra} and \texttt{PPPC4DMID}, and these two tools handle the electroweak corrections differently (see Ref.~\cite{Bauer:2020jay} for a discussion). The source energy spectra are different in the two codes and consequently this fact might impact the sensitivity of the detector.

Figure~\ref{fig:main_plot} shows expected sensitivities for the representative $b \bar{b}$, $\tau^+\tau^-$, and 
$\nu \bar{\nu}$ channels, respectively, as labelled. Our upper limits are confronted with the existing constraints available from the literature, obtained from different neutrino detectors. These include the combined limits from  IceCube~\cite{IceCube:2017rdn}, Antares~\cite{Albert:2016emp} and Super-Kamiokande~\cite{Mijakowski:2020qer}, see Ref.~\cite{Arguelles:2019ouk} for details.  For reference we also indicate the unitarity limit~\cite{Griest:1989wd,Smirnov:2019ngs} and the thermal relic cross-section~\cite{Steigman:2012nb}. Note here that we show the conservative unitarity limit of Ref.~\cite{Smirnov:2019ngs}, \emph{i.e.} the limit for composite self-conjugate dark matter. We do this to keep Fig.~\ref{fig:main_plot} model independent. We note that according to Ref.~\cite{Smirnov:2019ngs}, the unitarity limit for point-like DM is $140\,{\rm TeV}$.  
In general KM3NeT is expected to yield considerable improved upper limits as compared to other neutrino detectors. Importantly, we find that for the $\nu \bar{\nu}$ channel, both KM3NeT ORCA and ARCA may potentially reach the thermal relic cross-section. For the simplest WIMP hypothesis this cross-section represents the required cross-section to reproduce the observed DM abundance through the freeze-out mechanism. Thus, constraints on the \emph{total} DM cross section below the thermal limit can definitely test the WIMP hypothesis, as cross sections smaller than the thermal limit would over-close the Universe.  For many DM annihilation channels that involve electromagnetic particles (\emph{e.g.}, gamma rays and electrons/positrons), the exclusion limits are already probing the thermal limit and below it in some mass ranges, most particularly below $m_\chi \lesssim 10$ GeV~\cite{Leane:2018kjk}. The constraints on the \emph{total} cross section are often limited by channels with weak electromagnetic contributions, and especially, the $\nu \bar{\nu}$ channel, see \emph{e.g.} Ref.~\cite{Leane:2018kjk} for a thorough discussion. 
The possibility of KM3NeT reaching the thermal limit cross section will therefore be a important milestone for testing the WIMP hypothesis. Additionally, the ORCA detector is still under construction, its performances are improving and will become publicly available in the future. All this will augment the sensitivity of ORCA to detect neutrinos coming from DM masses in the range 100 GeV to 500 GeV by closing the present gap in energy (see the plot) between ORCA and ARCA for the $\nu \bar{\nu}$ annihilation channel.

\subsection{Discussions on Uncertainties}
\label{sec:uncertainties}

Our projected sensitivities make use of results from the literature presenting the KM3NeT experiment specifications, \emph{i.e.} Refs.~\cite{ KM3NeT:2021ozk, KhanChowdhury:2021kce}. 
However, the final completed experiment will have most likely performances that deviate from its original design. It is instructive to see how our results are affected if some of the variables describing the detector sensitivity changes with respect to the reference values we considered here.  

\begin{figure}
    \centering
    \includegraphics[width=0.85\textwidth]{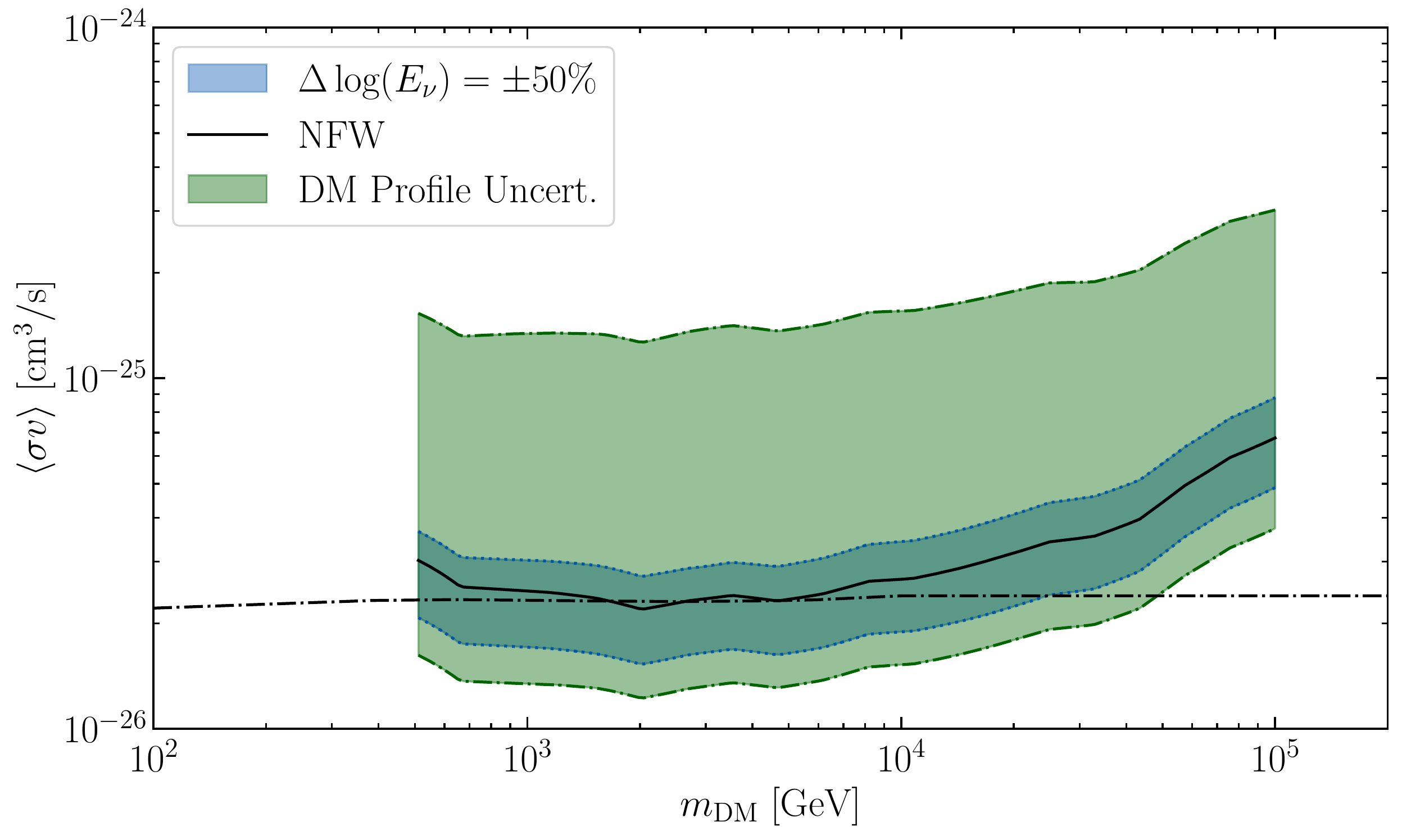}
    \caption{This panel shows the variation on the expected sensitivity reach of KM3NeT ARCA for the $\nu\bar{\nu}$ annihilation channel using NFW profile (solid line) and $50\%$ change in the energy resolution (dark green band) as benchmark. 
    For comparison, we also show the light green band bracketing the uncertainties related to the choice of the DM density profile: Burkert provides the upper dotted-dashed exclusion limit, while Einasto gives the lower dotted-dashed limit. All upper limits are at 95\% CL.}
    \label{fig:uncertainties}
\end{figure}

The most important parameter characterising the detector is the energy resolution. 
A well-known issue of neutrino telescope is the accurate energy reconstruction of the incoming neutrinos. Especially, high-energy muon flavoured neutrinos often lead to energetic long-muon tracks that can not be fully contained in the instrumented volume of the experiment.
Although KM3NeT, which is a water-Cherenkov telescope, is expected to perform better in this aspect than ice-Cherenkov telescopes, such as IceCube, this will be a major factor affecting the final performance for DM annihilation searches, especially for the ARCA detector.
To assess how variations in the energy resolution affect the sensitivity of the ARCA detector, we consider the benchmark scenarios that the energy resolution is better or worse than our default value of $\Delta \log_{10}(E_{\nu}) = 0.27$. We find that if the energy resolution is improved or worsened by $ 15\%$ (or $ 30\%$), the upper limits on the neutrino lines are improved about $ 10\%$ ($ 20 \%$) or worsened by $ 5\%$ ($ 10\%$). In case we consider a maximal variation of $ 50\%$, the gain or loose in sensitivity is around $ 30\% $ as shown in Fig.~\ref{fig:uncertainties}. As expected, the $\nu \bar{\nu}$ annihilation channel is the most correlated with changes in the energy resolution due to the shape of the signal featuring a narrow peak, while changes in the energy resolution affect much less the final states that feature a continuum energy spectrum. 

In case of ORCA, the flavour identification will significantly play a role in the sensitivities. It is expected that the use of each flavour will enhance the signal-to-background separation and improve the sensitivity. Due to computational limitations and complexity, this improvement, together with ameliorated energy and angular resolutions, is not studied here and left for future studies. 

For DM annihilation, KM3NeT is sensitive to the Galactic Centre region, where the expected signal is proportional to the DM density squared, see Eq.~\eqref{eq:dm_flux}.  Our results are reported using the NFW density profile Eq.~\eqref{eq:NFW}, which is widely used for many DM studies; this allows for a straightforward comparison with the results from the literature. The inner DM density profile scales with $r^{-1}$ for the NFW profile.  However, there are large uncertainties associated with the DM density shape in the inner galaxy; steeper or shallower DM density profiles are also viable choices for that region. To investigate the impact of the density profile choice on our results, we repeat the analysis by considering also the Burkert and the Einasto DM density profiles, with the shape parameters from Ref.~\cite{Cirelli:2010xx}. We find that the former leads to upper limits which are less stringent by a factor of $\sim 5$, while the latter produces stronger upper bounds by a factor of $\sim 2$, as illustrated in Fig.~\ref{fig:uncertainties}. Although choice of DM density profile is the source of the largest uncertainty, detector performance also plays a role for testing the WIMP hypothesis.

The shape of the DM density profile is actually the largest uncertainty in the theoretical prediction for the DM neutrino flux; as already said the uncertainties on the neutrino mixing parameters are negligible. The expected atmospheric neutrino background is also affected by uncertainties. In our background estimation for ORCA and ARCA we neglect the prompt component which is often the charm meson decays that is theoretically uncertain and remains undetected. In case of ARCA energy range, the prompt neutrino contribution is well below the astrophysical background below 100 TeV. As we stop at the 100 TeV energy and only use the track event topologies we do not include such background contribution, instead only include the astrophysical background.

\section{Implications for simple dark matter models}
\label{sec:models}

In this section we interpret the model independent results presented in the previous section for some simple DM models that will be particularly well probed by neutrino telescopes. We largely follow on the discussion in Ref.~\cite{BasegmezDuPree:2021fpo}, where the most promising models that only require one new mediator were highlighted. There the reach of a KM3NeT-like detector using a APS analysis was shown, for the case of heavy DM and ARCA only. APS analysis is robust under the choice of the DM density profile, but more conservative for DM profiles that are centrally peaked. Here we apply the the maximum-likelihood analysis performed above on a wide range of DM masses. 
The promising results shown in Fig.~\ref{fig:main_plot} where our sensitivity reaches the thermal relic line will be augmented by the fact that very few simple models can achieve 100\% branching ratio into neutrinos. 

DM interacting solely with neutrinos would of course be the most fortuitous scenario for neutrino telescopes. However, due to $SU(2)$ gauge invariance, there are only few simple models that completely evade constraints from $\gamma$-ray and direct DM detection experiments~\cite{Olivares-DelCampo:2017feq,Blennow:2019fhy,ElAisati:2017ppn}. Additionally, annihilation that is velocity independent ($s$-wave) is generally required for signals to be detectable at the next generation of neutrino telescopes. When considering DM as a Dirac Fermion, $\chi$, this is achieved for models with a scalar mediator, via $t$-channel (based on \emph{e.g.}~\cite{Arina:2020udz}) and models with a vector mediator, via $s$-channel annihilation. For the latter, we are going to discuss a gauged $U(1)_{L_{\mu}-L_{\tau}}$ model, as it is more theoretically interesting, especially in the light of recent anomalies in the muon sector~\cite{Bauer:2018onh, Muong-2:2021ojo} and a simple anomaly free model.

The model predictions have been computed using the \texttt{MadDM}~\cite{Ambrogi:2018jqj,Arina:2021gfn} tool. The gamma-ray bounds from the dwarf spheroidal galaxy measurements collected by the Fermi-LAT satellite~\cite{Fermi-LAT:2016uux} are computed using the same tools while direct detection exclusion limits and projections have been obtained following~\cite{Cerdeno:2018bty}. All details can be found in~\cite{BasegmezDuPree:2021fpo}, with the exception of the analysis of LZ, with their recently published in~\cite{LZ:2022ufs} and their future projection~\cite{LZ:2015kxe, Mount:2017qzi}. For their first data release~\cite{LZ:2022ufs}, after 60 live days and their 5.5t fiducial mass, the collaboration was able to put the most stringent constraints on spin-independent DM-nucleon interactions. We model this LZ limit similarly to Ref.~\cite{Chang:2022jgo}, by performing a 1-binned Poissonian analysis, where we can neglect any background by only considering events below the mean of the nuclear recoil band in the signal region, reducing the signal by 50\%. This method is unable to fully reproduce the limit obtained by LZ, but above $m_{\chi}\sim 200\,{\rm GeV}$ our result matches the official one in Ref.~\cite{LZ:2022ufs}. As will be seen below, this is sufficient for our purposes because the only time we consider $m_{\chi}\leq 200\,  {\rm GeV}$ is when direct detection constraints can be avoided completely. For the projections of LZ after its full data taking period, instead of following~\cite{LZ:2015kxe, Mount:2017qzi} we simply apply the same analysis as in Ref.~\cite{LZ:2022ufs} but scale up the exposure from $60$ to $1000$ days.

{\it The scalar mediated case:} This model introduces a new $SU(2)_{{\rm EW}}$ scalar doublet $\varphi$ that only couples simultaneously to the SM lepton doublets $L_{\alpha}$ ($\alpha=e,\mu,\tau$) and the DM $\chi$,
\begin{equation}
\mathcal{L}^{\varphi} = y_{\alpha} \bar{\chi} L_{\alpha} \varphi^{\dagger} + \rm h.c.\,, 
\label{eq:scalar_lag}
\end{equation}
where $y_\alpha$ coupling strength is a $3\times 3$ matrix in flavour space, that we take real for simplicity.  The new scalar field is also odd under the dark symmetry that stabilises the DM particle. The annihilation process is through the $t$-channel diagram exchanging the scalar mediator. In order for $\chi$ to be the DM particle it is required $m_{\chi}\leq m_{\varphi}$.
\begin{figure}[t]
    \includegraphics[width=0.5\textwidth]{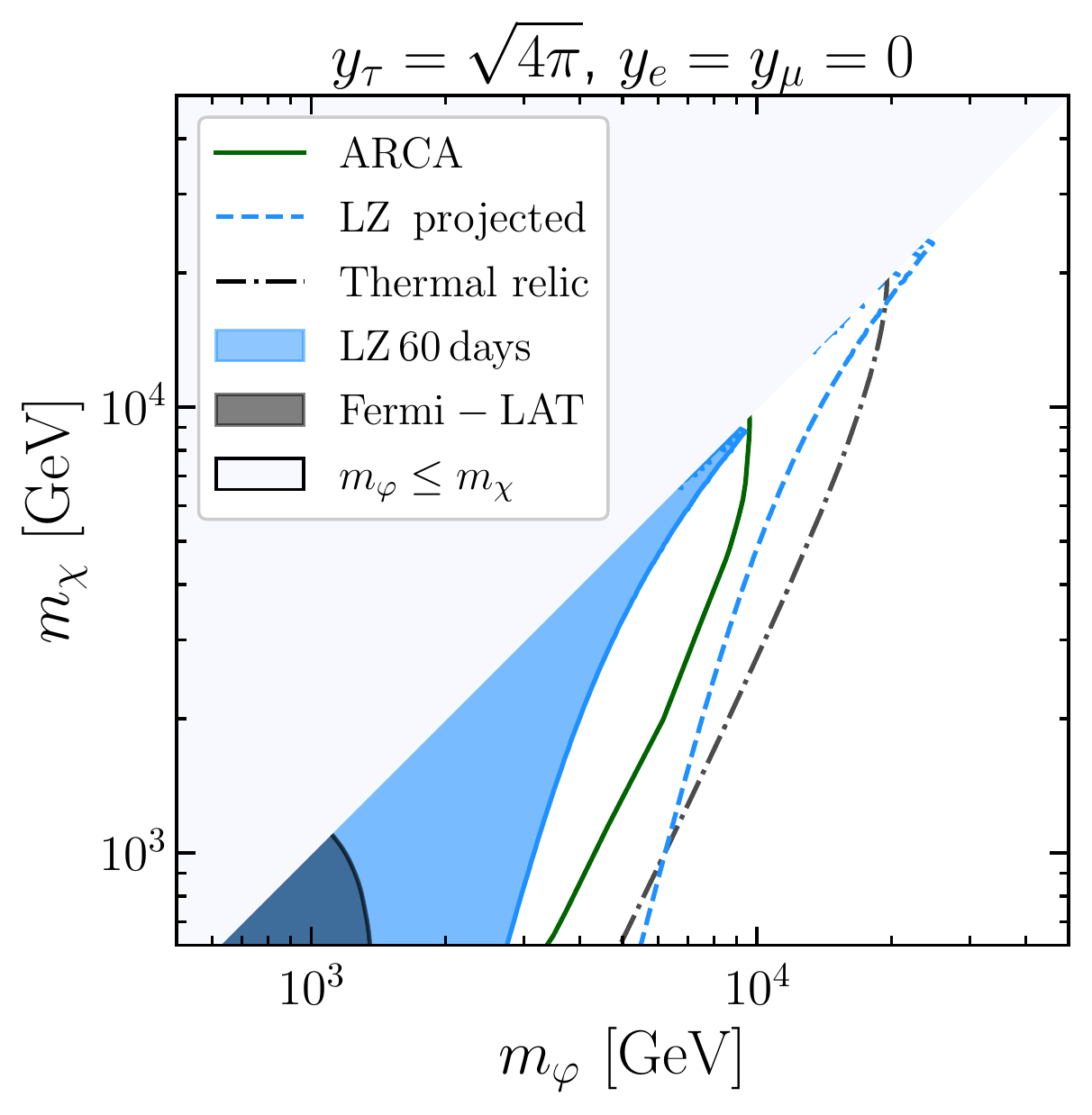}
    \includegraphics[width=0.5\textwidth]{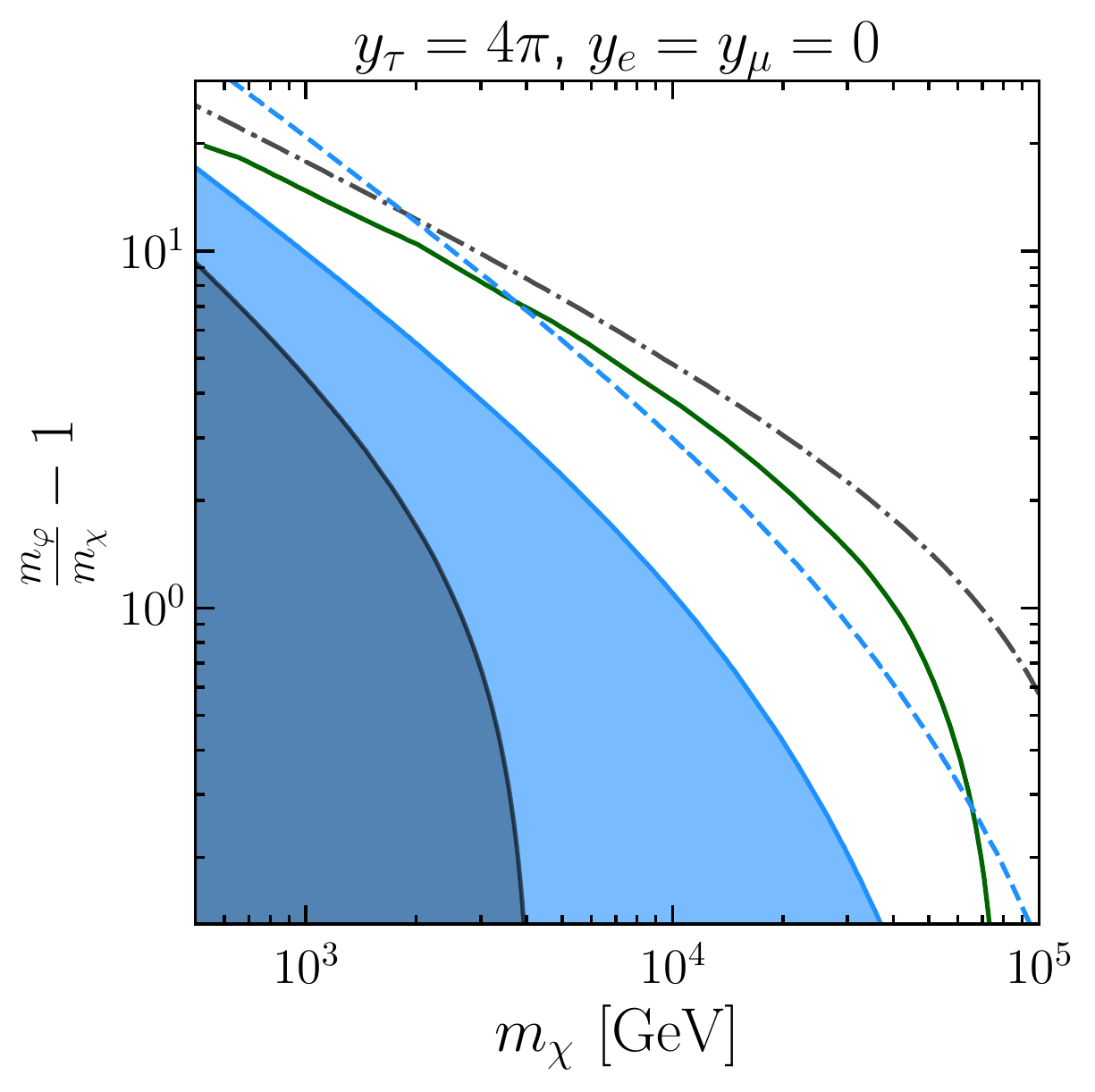}
\caption{\textbf{Left:} ARCA projections (solid green) for the $t$-channel model compared with the existing and future LZ (light blue shaded and dashed blue respectively) and Fermi-LAT (black shaded) limits, in the $\{ m_{\chi}, m_{\varphi} \}$ mass-plane. The coupling strength is fixed at $y_{\tau}=\sqrt{4 \pi}$. The dash-dotted line denotes the relic density line. Inside the line the DM is under-abundant, while outside the DM is over-abundant. All exclusion limits and projections are provided at 95\% CL. \textbf{Right:} Same as left for a coupling strength fixed at $y_{\tau}=4\pi$ in the $\{ (m_\varphi/m_\chi-1),  m_{\chi} \}$-plane.}
    \label{fig:tchan_lims}
\end{figure}

Evidently, the cross-sections for $\chi\bar{\chi}\rightarrow\bar{\nu}\nu$ and $\chi\bar{\chi}\rightarrow\ell^+\ell^-$ are equivalent when $m_{\chi} > m_{\ell}$. This then means that $\gamma$-ray constraints will compete with the projections for KM3NeT-like detectors. Additionally, as pointed out in~\cite{Kopp:2014tsa,Ibarra:2015fqa}, loop processes introduce DM nucleon interactions which could be observed in direct detection experiments. The already very high sensitivities of both $\gamma$-ray and direct detection experiments in the low DM mass region below and around 100 GeV mean that ORCA does not promise to provide new insights into this model. However, our projections for ARCA, as shown in Fig.~\ref{fig:tchan_lims}, will be competitive.  

In this figure we show how the ARCA projections compare to exclusion limits for a continuum gamma-ray spectrum provided by the Fermi-LAT bounds and the LZ upper bound on the nuclear recoil rate. We see that ARCA can provide a substantial improvement on the current limits for such a model. The key feature which plays into ARCA's favor is its sensitivity at high energy, while direct detection and $\gamma$-ray experiments being most sensitive in the $m_{\chi}\sim 100\,{\rm GeV}$ and $m_{\chi}\sim 10\,{\rm GeV}$ realms respectively. It would be remiss to not mention the timescales at work here, the LZ full exposure will be reached years prior to the KM3NeT limit we have show.   

{\it A gauged $U(1)_{L_{\mu}-L_{\tau}}$:} as outlined in~\cite{BasegmezDuPree:2021fpo}, this model is particularly appealing for a phenomenological study, as it is a simple extension to the SM which does not possess any theoretical ailments such as gauge anomalies and the related unitarity violation. Furthermore, for certain masses of the new vector $Z'$, it can lead to a DM annihilation branching ratio to neutrinos of 100\%. The Lagrangian reads as follows,
\begin{align}\label{eq:schann}
 {\cal L} = g_{\chi}\bar\chi \gamma_{\alpha}
  \chi\,Z'^{\alpha}  + g_{\mu-\tau}\left(\bar \mu_R \gamma_{\alpha} \mu_R Z'^{\alpha} -\bar \tau_R \gamma_{\alpha} \tau_R Z'^{\alpha}+ \bar L_{\mu} \gamma_{\alpha} L_{\mu} Z'^{\alpha} -L_{\tau} \gamma_{\alpha} L_{\tau} Z'^{\alpha}\right)
    \,,
\end{align}
where $\mu_R, \tau_R$ should be understood as the right-handed muon/tau leptons and $L_{\mu,\tau}$ as the leptonic $SU(2)$ doublet of the muon and tau generation. We have written here $g_{\chi}$ and $g_{\mu-\tau}$ as separate couplings, but in analogy with the SM, one may think it is more natural to take a universal gauge coupling and express any difference in terms of charges under the $U(1)_{L_{\mu}-L_{\tau}}$ group. However the distinction between $g_{\chi}$ and $g_{\mu-\tau}$ is convenient to illustrate two different regimes: (i) $g_{\chi} \sim g_{\mu-\tau} \sim 1$ and (ii)  $g_{\chi} >> g_{\mu-\tau}$ that is called secluded DM~\cite{Pospelov:2007mp}. 

\begin{figure}
    \includegraphics[width=0.5\textwidth]{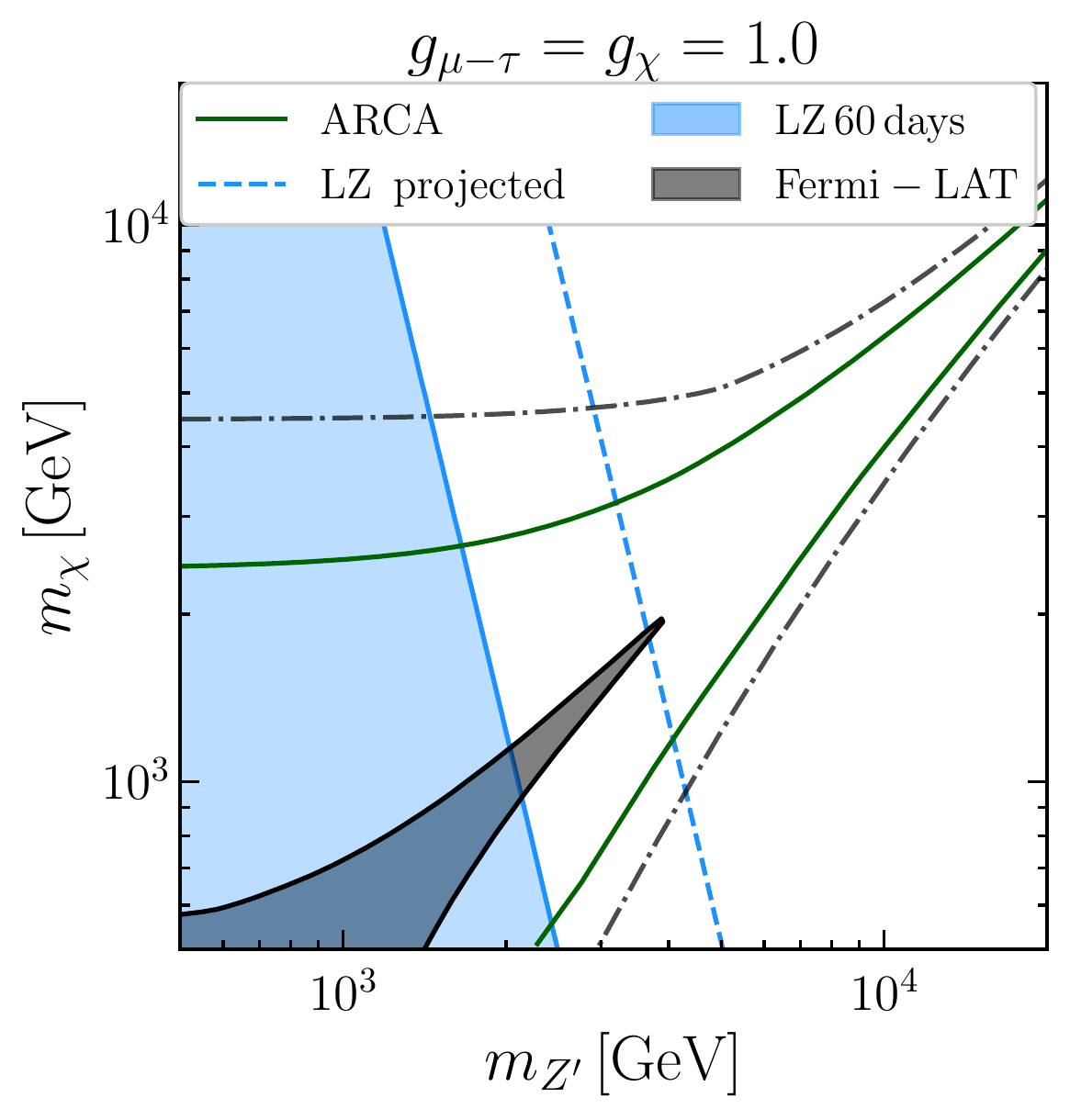}
    \includegraphics[width=0.5\textwidth]{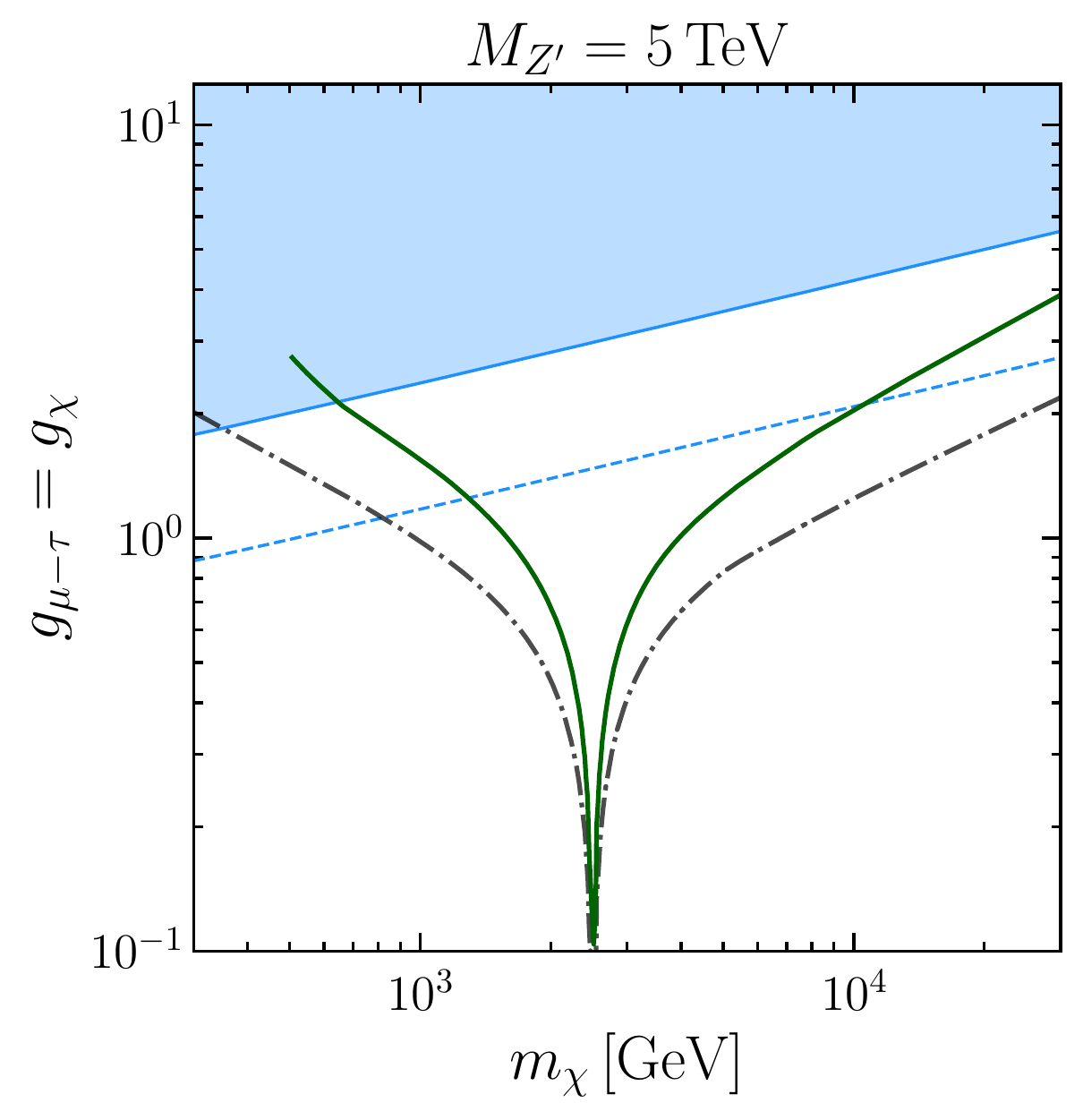}
    \caption{Left: Projection for the ARCA detector (solid green) for the gauged $U(1)_{L_{\mu}-L_{\tau}}$ model in the $\{ m_{\chi}, m_{Z' } \}$ mass-plane for fixed couplings $g_\chi = g_{\mu-\tau}=1$. The reach of ARCA is compared with the existing XENON1T (light blue shaded region) and Fermi-LAT (gray region) limits. The dash-dotted line denotes the relic density line. Inside the line the DM is under-abundant, while outside the DM is over-abundant. All exclusion limits and projections are provided at 90\% CL. Right: Same as left in the $\{ m_{\chi}, g_\chi=g_{\mu-\tau} \}$ plane for a fixed $Z'$ mass of 5 TeV.}
    \label{fig:gauged_s_chan}
\end{figure}

Considering first the case (i), DM annihilation into leptons proceeds through $s$-channel mediated by a relatively heavy $Z'$ ($m_{Z'}$ and $m_{\chi}$ have similar orders of magnitude). There is no natural way here to suppress annihilation into charged leptons with respect to neutrinos, once again gamma-ray and direct detection bounds will be the most sensitive probes for this model for $m_\chi \lesssim 100$ GeV. On the other hand, when the DM mass increases considerably above the TeV, Fermi-LAT and LZ loose sensitivity and neutrino telescope become a unique probe for leptophilic models. The projections for ARCA are shown in Fig.~\ref{fig:gauged_s_chan} for different slices of the model parameter space.
We see that ARCA will be able to investigate a large part of unexplored parameter space, which direct and gamma-ray experiments will not be able to access in a near future. Remarkably ARCA will get close to the thermal relic line in case of resonant annihilation, as shown in the right panel for a benchmark $Z'$ mass of 5 TeV. This demonstrates the complementarity of neutrino telescopes with LHC experiments that are unable yet to probe new vector bosons at such large masses~\cite{CMS:2021ctt, ATLAS:2019erb}.

In regime (ii), secluded DM features a very light mediator ($m_{Z'} \ll m_\chi$), perturbative but large $g_\chi$ while negligible coupling to the SM sector, $g_{\mu-\tau} \lesssim 10^{-4}$ at least. For such coupling strength hierarchy, the dominant annihilation channel is $\chi \bar{\chi} \to Z' Z'$, which implies that the DM  relic density value is set solely by $g_\chi$. Even though $g_{\mu-\tau}$ is extremely tiny, the $Z'$ is not stable and will eventually decay into four SM leptons. Annihilation into charged leptons, when kinematically allowed, takes place with equal branching ratios and gives rise to a $4\mu$ or $4\tau$ final state. The light $Z'$ mediators are always boosted with respect to the DM annihilation reference frame. Subsequently, they decay into 4 neutrinos with box-shaped energy spectra, see \emph{e.g.} Refs.~\cite{Ibarra:2012dw,Garcia-Cely:2016pse,ElAisati:2017ppn}. 

In this secluded regime the direct DM constraints are quite prohibitive on the size of $g_{\mu-\tau}$. One can see in Ref.~\cite{Cheek:2021grh} that even Xenon1T~\cite{Aprile:2018dbl} can constrain values $g_{\mu-\tau}\gtrsim 10^{-9}$ for a given DM mass and $g_{\chi}$. This is because the mediator is no longer suppressing the nuclear recoil rate. Since the $U(1)_{L_{\mu}-L_{\tau}}$ model is leptophilic at tree level, it is conceivable that the kinetic mixing $\epsilon$, is in fact zero~\cite{Heeck:2022znj}, thus completely avoiding direct detection with nuclear recoils. In such a case it could be possible to explain the $(g-2)_{\mu}$ anomaly with $m_{\chi}\gtrsim 5\,{\rm GeV}$. To us however, it is unclear what mechanisms exist to fully cancel out the loop-level contributions to the kinetic mixing term. Therefore the most promising scenarios for this model to account for both the measured values of the $(g-2)_{\mu}$ and DM is ones where DM is rather light $m_{\chi}< 1\,{\rm GeV}$~\cite{Foldenauer:2018zrz, Holst:2021lzm}, which is rather incompatible for the searches performed by KM3NeT. 

In such situations where the $g_{\mu-\tau}$ coupling is so small, there are a number of cosmological considerations which could constrain our model further, or even alter the expectation that DM is produced via thermal freeze-out. In our previous work~\cite{BasegmezDuPree:2021fpo} we conservatively took the constraint that the $Z^{\prime}$ must decay quickly enough as to not effect Big Bang Nucleosynthesis. This roughly translates to $\tau_{Z^{\prime}} < 1\,{\rm s}$~\cite{Kawasaki:2017bqm}. It turns out, that below the di-muon threshold, decays are overwhelmingly into neutrinos, and contributes to heating of the neutrino temperature~\cite{Escudero:2019gzq,Aghanim:2018eyx} and consequently, contributes to the effective relativistic degree of freedom at recombination $\Delta N_{\rm eff}$. Of course, the $Z^\prime\rightarrow e^+e^-$ decay occurs, but it is loop-suppressed, and for low values of $g_{\mu-\tau}$, the kinetic mixing is even smaller, relevant constraints can be found in Ref.~\cite{Fradette:2014sza}. Indeed from this one can obtain lower bounds on $g_{\mu-\tau}\geq 10^{-11}$. However care is required, because the constraints of Ref.~\cite{Fradette:2014sza} rely only on the kinetic mixing interaction to produce the relic yield of $Z^\prime$, which would not be true in our case. This leads us to another aspect of the phenomenology that requires some care. As couplings to the SM decrease, assumptions about $Z^\prime$ and $\chi$ being in thermal equilibrium may no longer be valid. Indeed, according to Ref.~\cite{Escudero:2019gzq}, in order for the $Z^\prime$ to achieve equilibrium with the SM thermal bath one requires $g_{\mu-\tau}\gtrsim4\times 10^{-9}$. 

What cannot be avoided are bounds coming the Planck~\cite{Aghanim:2018eyx} satellite on the recombination epoch. Indeed $\langle \sigma v \rangle$ is severely boosted by the small velocities at that epoch, as a light mediator induces Sommerfeld enhancement, which we have properly included as in~\cite{Hisano:2004ds,ArkaniHamed:2008qn,Iengo:2009ni,Cassel:2009wt,Arina:2010wv}. 
\begin{figure}
    \includegraphics[width=0.5\textwidth]{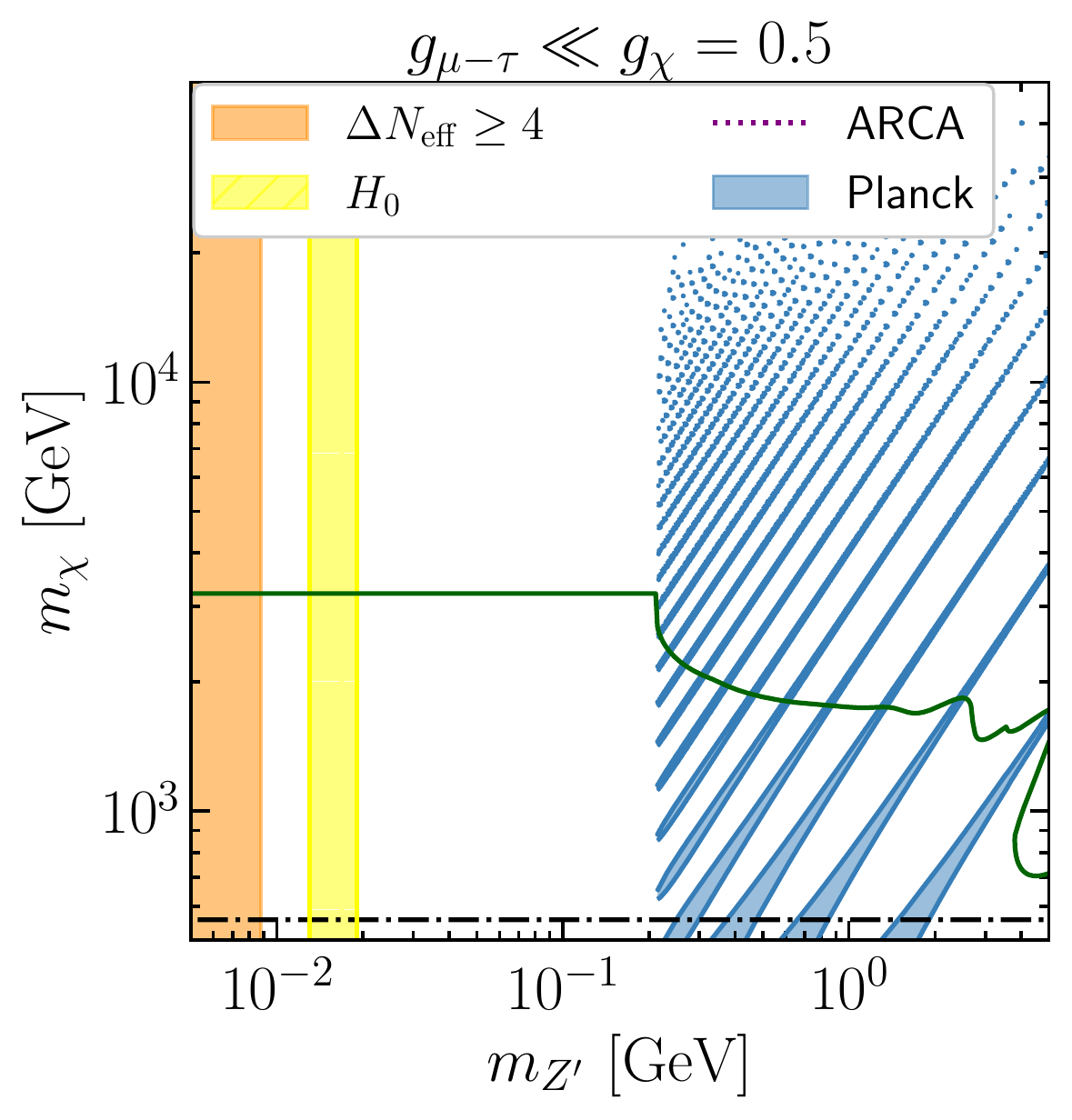}
    \includegraphics[width=0.5\textwidth]{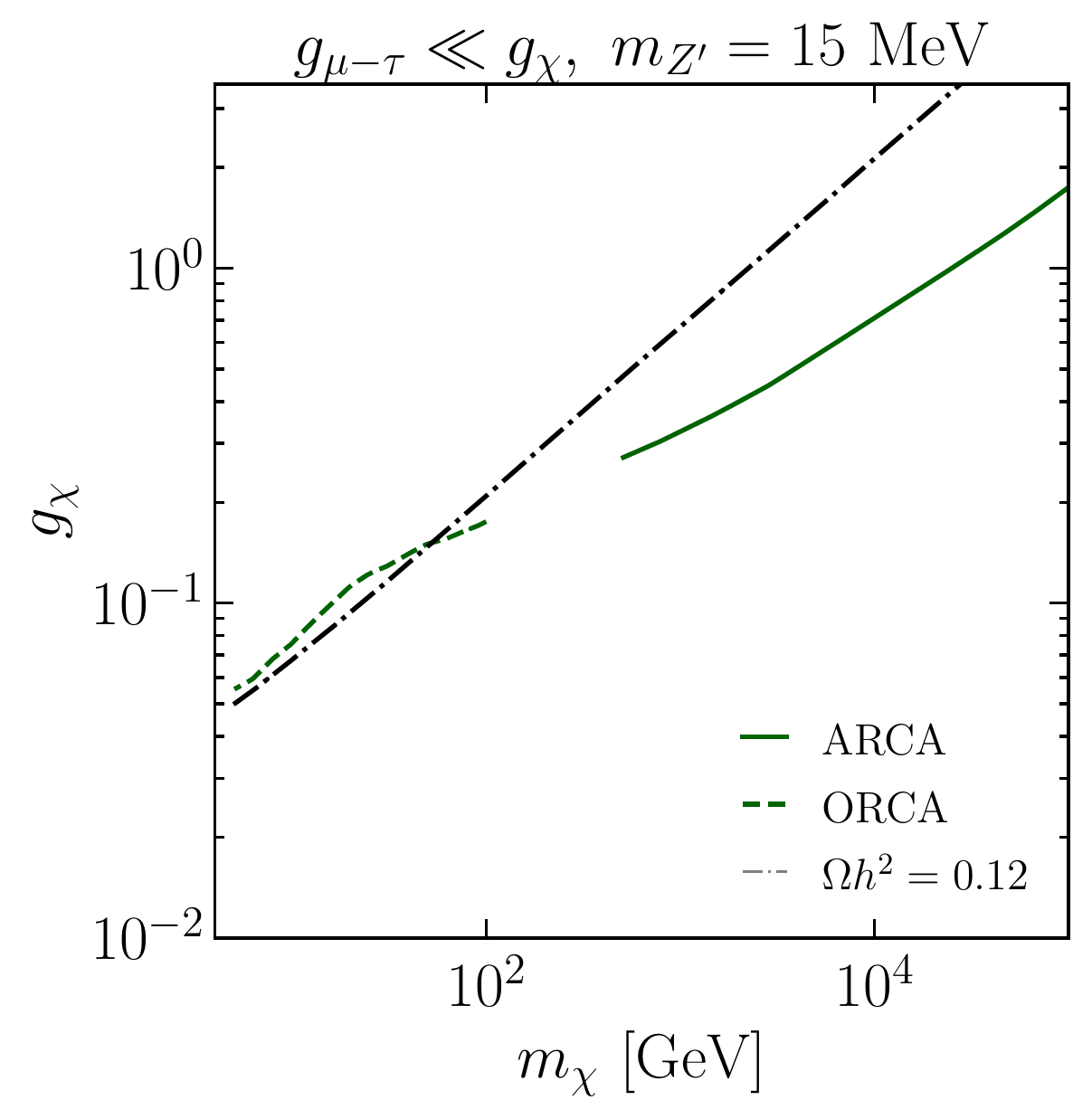}
    \caption{Left: Projection for the ARCA detector (solid green) for the gauged $U(1)_{L_{\mu}-L_{\tau}}$ model in the secluded scenario. The light blue shaded region is disfavored by current Planck exclusion limit. while the orange region is disfavoured by self-interactions and measurements of $\Delta N_{\rm eff}$. The yellow region denotes the model parameter space favoured to alleviate the Hubble tension. The dashed line indicates the relic density line. Right: Projection for both ORCA and ARCA detectors (solid green) for the gauged $U(1)_{L_{\mu}-L_{\tau}}$ model in the secluded scenario in a wide range of DM masses as a function of $g_{\chi}$. The dashed line indicates the relic density line.}
    \label{fig:gauged_secluded}
\end{figure}

The result for the ARCA detector is shown in the left panel of Fig.~\ref{fig:gauged_secluded}. Below the muon mass threshold, the $Z'$ can only decay into neutrinos making ARCA a unique probe for secluded thermal DM in the very light $Z'$ mass range.  There the only bound on the model comes from measurements of the effective relativistic degree of freedom before recombination~\cite{Escudero:2019gzq,Aghanim:2018eyx} ($\Delta N_{\rm eff}$, red shaded region), which sets a bound on the $Z'$ mass to be larger than 4 MeV roughly, by asking $\Delta N_{\rm eff} \leq 4$. Self-interaction constrains the size of the scattering process $\chi \chi \to \chi \chi$, see \emph{e.g.},~\cite{Bringmann:2016din}, impacting a region very similar to the $\Delta N_{\rm eff} \leq 4$ one, hence are not shown. For $m_{Z'}\sim 10 $ MeV, such light vector boson can also contribute to the resolution of the Hubble tension~\cite{Aghanim:2018eyx,Riess:2016jrr,Riess:2018byc,Riess:2019cxk} (yellow shaded band) by increasing $\Delta N_{\rm eff}$ up to roughly $0.2-0.4$~\cite{Bernal:2016gxb,Escudero:2019gzq}. This value is required to reconcile the determination of $H_0$ from local measurements with the one from the cosmic microwave background. As soon as DM annihilation into charged leptons is kinematically allowed, bounds from Planck already exclude much of the model parameter space, although in a sporadic way, due to the resonant enhancement from the Sommerfeld effect.

In Fig.~\ref{fig:gauged_secluded} (right panel) we show the reach of both ORCA and ARCA detectors for this specific situation where annihilation is almost exclusively into neutrinos, in the $m_{\chi},\,g_{\chi}$ plane. We see that with the aid of the Sommerfeld enhancement, ARCA will be able to constrain this model where thermal freeze-out would produce the correct relic abundance. Additionally, for the first time in this section, we see that ORCA will be able to provide leading bounds on a model we consider. Due to the scale separation of $m_{\chi}$ and $m_{Z'}$ we see the Sommerfeld effect even contributing to the ORCA projections, bringing the sensitivity to the thermal line as $m_\chi\rightarrow100\,{\rm GeV}$ despite the decrease in sensitivity shown for a similar region in the model independent limits of Fig.~\ref{fig:main_plot}.

\section{Conclusions}
\label{sec:concl}

In this study we have explored the sensitivity of a fully-constructed KM3NeT experiment to DM annihilation by means of its low-energy ORCA and high-energy ARCA detectors. 
DM annihilation signals correlate with the Galactic Centre, hence we have based our analysis on the most central region of the Milky Way and developed a likelihood-ratio analysis using energy and angular variables of both track and shower events. We have taken into account the expected performance of the experiment as detailed in the LoI~\cite{KM3Net:2016zxf} and optimised the analysis method accordingly. We have enhanced the sensitivity for ORCA by combining track and shower events, while we have only used track topologies for ARCA because of its superior angular resolution. The signal and background events have been generated according to their theoretical expectations and/or measured data. As for the DM signal, we have taken as baseline the NFW density profile to describe the DM distribution in the galactic halo. It should be noted that the limited knowledge of the DM density profile in the central galactic region is by far the largest source of theoretical uncertainty for the forecasts, neutrino oscillation and mass parameters choices play a smaller role. We have quantified the variation in the projected sensitivities according to different choices of DM density profile, which is decreased by a factor of 5 or increased by a factor of 2 for a shallower or steeper density profile respectively. 
As for the background signal, we have used the atmospheric neutrino events for ORCA and fully incorporated the effects of oscillation in matter that neutrinos experience while travelling through Earth, which have an impact for neutrinos below 20 or 50 GeV depending on the flavour. For ARCA, atmospheric neutrinos are the main source of background up to energies of around 100 TeV, after which astrophysical neutrinos start to dominate the background signals. We have then obtained the model-independent upper limits on the thermally averaged annihilation cross-section in a wide range of DM masses, $ 5 \, \rm GeV \lesssim m_\chi \lesssim 10^5 \, \rm GeV$, and for the most optimal channels into 2 or 4 leptons, including monochromatic neutrino lines and boxes. As shown in Fig.~\ref{fig:main_plot}, we have found that neutrino experiments like KM3NeT will be able to probe large portions of the WIMP parameter space in the next decade by reaching the freeze-out values for $\langle \sigma v\rangle$ for DM masses around the TeV scale for annihilation into pure neutrino channels.

Although the experimental setup we used here is determined by the design specifications with limited knowledge and performance of the actual experiment, the real experiment might be different and outperform our projections. We have investigated such improvements in Sec.~\ref{sec:uncertainties}: a $15\%$ gain in the performance of track event topologies will reflect in a $ 10\%$ enhancement in the sensitivity. So far shower events can not be easily incorporated due to the lack of (public) knowledge about the performance of the ARCA detector to such events. We expect that the inclusion of the latter events and the use of both event topologies combined will further improve the sensitivity for ARCA. With possible amelioration in the neutrino detection and flavour identification techniques, analysis methods and detector upgrades, we expect substantial gain in the DM detection with neutrinos in the near future. 

We then have pursued a phenomenological interpretation of our sensitivities considering DM models that give rise to observable neutrino signal in the future KM3NeT-like telescope. We have selected few simple and motivated leptophilic models in which Dirac DM particles annihilate into neutrinos via $s$-wave. We have found that for the $U(1)_{L_\mu-L_\tau}$ model, ARCA will be sensitive to the thermal freeze-out part of the parameter space for DM masses above the TeV. This is an important milestone for neutrino telescopes, as it denotes the capability to test the WIMP hypothesis in a mass range, \emph{e.g.} around $1$ to $10$ TeV, which is often at too high-energy for other DM probes and for LHC to be very sensitive. 

Our assumptions for the modelling part are not restrictive requirements as: (i) there are plenty of realisation that involve also other types of DM particles, see \emph{e.g.}~\cite{ElAisati:2017ppn} and our previous work~\cite{BasegmezDuPree:2021fpo}, (ii) annihilation into charged leptons is presented because of $SU(2)$ gauge invariance and electroweak corrections from high-energy neutrinos produce nevertheless photon signals detectable by gamma-ray telescopes, and (iii) one-loop corrections typically induce a non-zero DM-nucleon scattering cross-section accessible to direct detection experiments. Hence the theoretical scenarios discussed in this paper provide a good starting point for the comparison of different probes. As shown, there is a large parameter space that can be reached by the neutrino telescopes. This provides a crucial information to other search experiments. In case of a signal is observed, the models investigated in this paper will provide a starting point to reveal the nature and type of interactions of DM particles. 

\acknowledgments
We would like to thank Shin'ichiro Ando for discussions, Thijs Juan van Eeden and Jordan Seneca for clarification about the KM3NeT experiment performance, Soebur Razzaque for discussions about neutrino oscillations, Fabio Maltoni, Filippo Sala and Darren Scott for discussions concerning the secluded $U(1)_{L_{\mu}-L_{\tau}}$ model and Ibles Olcina Samblas, Felix Kahlhoefer, Darren Price and Ellen Sandford for discussions on the LZ first result. 

SB is supported by NWO (Dutch National Science Organisation) under project no: 680-91-004. AC is supported by the grant ``AstroCeNT: Particle Astrophysics Science and Technology Centre" carried out within the International Research Agendas programme of the Foundation for Polish Science financed by the European Union under the European Regional Development Fund.

\bibliographystyle{jhep}
\bibliography{bibliographyfile}

\end{document}